# Puzzles in Hadronic Physics and Novel Quantum Chromodynamics Phenomenology

Stanley J. Brodsky, (1) Guy de Téramond,(2) and Marek Karliner (3)

(1) SLAC National Accelerator Laboratory, Stanford University, Stanford, CA 94309, USA
(2) National Academy of Sciences, Universidad de Costa Rica, 2060 San José, Costa Rica
(3) Raymond and Beverly Sackler School of Physics and Astronomy,
    Tel Aviv University, Tel Aviv 69978, Israel

**Abstract**

We review some outstanding puzzles and experimental anomalies in hadron physics that appear to challenge conventional wisdom and, in some cases, the foundations of quantum chromodynamics (QCD). We also discuss possible solutions and propose new tests and experiments that could illuminate the underlying physics and novel phenomenological features of QCD. In some cases, new perspectives for QCD physics have emerged.

*Published in Ann. Rev. Nucl. Part. Sci. 2012, 62 (2012).*

Work funded in part by DOE Contract DE-AC02-76SF00515

# Contents

# 1. INTRODUCTION

Quantum chromodynamics (QCD) is believed to be the underlying theory of hadron and nuclear physics. It provides a fundamental description of the structure of hadrons and nuclei in terms of elementary quark and gluon fields, just as quantum electrodynamics (QED) provides a fundamental theory of electromagnetism, atomic physics, and chemistry. The most challenging problem of strong interaction dynamics remains how to determine the composition of hadrons in terms of their fundamental QCD quark and gluon degrees of freedom. Extensive empirical evidence validates QCD; nevertheless, numerous phenomena observed in experiments appear to contradict our QCD-based intuition about hadron and nuclear physics.

We review some outstanding puzzles and experimental anomalies in hadron physics that appear to challenge conventional wisdom and, in some cases, the foundations of QCD. We also discuss possible solutions and propose new tests and experiments that can illuminate the underlying physics and novel phenomenological features of QCD. In some cases, surprising new perspectives for QCD physics have emerged.

# 2. THE $J/\psi \to \rho\pi$ PUZZLE

The largest observed two-body hadronic decay rate of the $J/\psi$ $c\bar{c}$ 1S resonance is $J/\psi$ to a $\rho$ and a pion: $\Gamma(J/\psi \to \rho\pi)/\Gamma_{\text{tot}} = (1.69 \pm 0.15) \times 10^{-2}$. However, the $\psi(2S)$, which is the radial excitation of the $J/\psi(1S)$, has a very small branching ratio to this channel, $\Gamma[\psi(2S) \to \rho\pi]/\Gamma_{\text{tot}} = (3.2 \pm 1.2) \times 10^{-5}$ (1):

$$\frac{\Gamma(\psi(2S) \to \rho\pi))/\Gamma_{\text{tot}}}{\Gamma(J/\psi \to \rho\pi))/\Gamma_{\text{tot}}} = \frac{3.2 \pm 1.2 \times 10^{-5}}{1.69 \pm 0.15 \times 10^{-2}} \simeq 0.2\%. \qquad 1.$$

If quarkonium decay occurs via $c\bar{c} \to ggg \to \rho\pi$, then the observed ratio of branching ratios for $\psi(2S)$ and $J/\psi$ decay for every channel should track with the wave function at the origin and, thus, have the same ratio of branching ratios (13%) as the decays to lepton pairs:

$$\frac{\Gamma(\psi(2S) \to e^+e^-)/\Gamma_{\text{tot}}}{\Gamma(J/\psi \to e^+e^-)/\Gamma_{\text{tot}}} = \frac{7.73 \pm 0.17 \times 10^{-2}}{5.94 \pm 0.06 \times 10^{-3}} \simeq 13\%. \qquad 2.$$

This two-orders-of-magnitude discrepancy in the ratio of $J/\psi(1S)$ and $\psi(2S)$ branching ratios for $\rho\pi$, versus the lepton-pair branching ratios, contradicts the OZI (Okubo–Zweig–Iizuka) rule and the standard assumption in QCD that a heavy quarkonium bound state decays to hadronic final states at short distances through the annihilation of the valence $c$ and $\bar{c}$ to three gluons, each of which has virtuality $q^2 \simeq M_\psi^2/9$ (**Figure 1a**). This puzzle is compounded by the fact that all vector-pseudoscalar meson decay channels of the charmonium states should be strongly suppressed due to a perturbative QCD (pQCD) principle: hadron helicity conservation (2).

Note that because of parity conservation, the $\rho$ in $\psi \to \rho\pi$ is always produced with polarization $J^z = \pm 1$. Thus, either the decay to $\rho\pi$ requires a helicity flip of the light quark or it must be produced via nonzero orbital angular momentum $L^z = \pm 1$; both mechanisms are expected to be strongly suppressed. The $\psi(2S)$ decays obey hadron helicity conservation, but this pQCD rule is severely violated by the large rate for $J/\psi \to \rho\pi$.

This puzzle could be resolved (3) if $J/\psi \to \rho\pi$ occurs by a mechanism in which the heavy quark pair never annihilates but instead propagates to the higher Fock states of the final-state mesons, which contain intrinsic $c\bar{c}$ pairs; in other words, the charm quarks do not annihilate into gluons but simply flow into the $|q\bar{q}c\bar{c}\rangle$ Fock state of the final-state $\rho$ or $\pi$ (**Figure 1b**). An essential point is that this mechanism would be strongly suppressed for $\psi(2S) \to \rho\pi$ because the wave function of the $\psi(2s)$ has a radial node that severely decreases the strength of the matrix

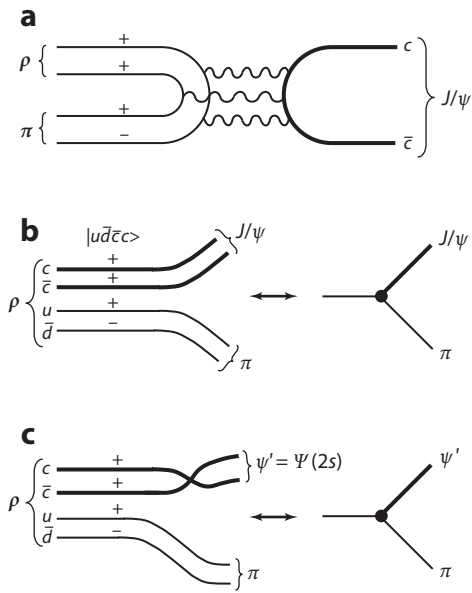

**Figure 1**

$J/\psi$ decay to $\rho\pi$. (*a*) The standard pQCD (perturbative quantum chromodynamics) three-gluon exchange amplitude is suppressed by the need for a quark helicity flip. (*b*) Coupling of the $J/\psi$ directly to the intrinsic charm Fock state of the $\rho$ meson. (*c*) Suppression of the intrinsic charm mechanism due to the node in the $\rho'$ wave function.

element evaluated from the spatial overlap of quarkonium and final-state meson wave functions (**Figure 1***c*). Equivalently, one can postulate that the hadronic decays of the $J/\psi$ and $\psi(2S)$ are mediated by $D\bar{D}$ intermediate states (4, 5).

The $J/\psi \to \rho\pi$ puzzle thus indicates a possible conflict with one of the basic premises of QCD: that the strong interaction is propagated by spin-one color-octet gluon degrees of freedom. In fact, in the Isgur–Paton model (6), gluonic interactions are replaced by the propagation of a flux tube. Similarly, in the anti–de Sitter (AdS)/QCD model, the interaction of gluons below virtuality $Q^2 \simeq 1\,\text{GeV}^2$ are sublimated in favor of an effective color-confining potential (7). The latter hypothesis can explain other puzzles in hadron physics, as discussed below.

## 3. THE ANOMALOUS $A_{NN}$ SPIN-SPIN CORRELATION AND COLOR TRANSPARENCY ANOMALIES IN PROTON-PROTON ELASTIC SCATTERING

The term transversity in hadron physics encompasses the entire range of spin, orbital angular momentum, and transverse momentum measures of hadron structure, which are accessible by experiment. Highly sensitive experiments such as HERMES at DESY (8), COMPASS (9–11) at CERN, and CLAS at Jefferson National Accelerator Laboratory (JLab) (12, 13) are now providing an extensive range of experimental results that, in turn, are providing new insights into the fundamental quark and gluon structure of the nucleons. The challenge for theory is to synthesize this information into a consistent picture of hadron dynamics and to confront QCD at a fundamental level. The historic example of transversity is the remarkably large spin-spin correlation in $pp$ elastic scattering measured by Krisch and collaborators (14) in which the beam and target

*4   Brodsky • de Téramond • Karliner*

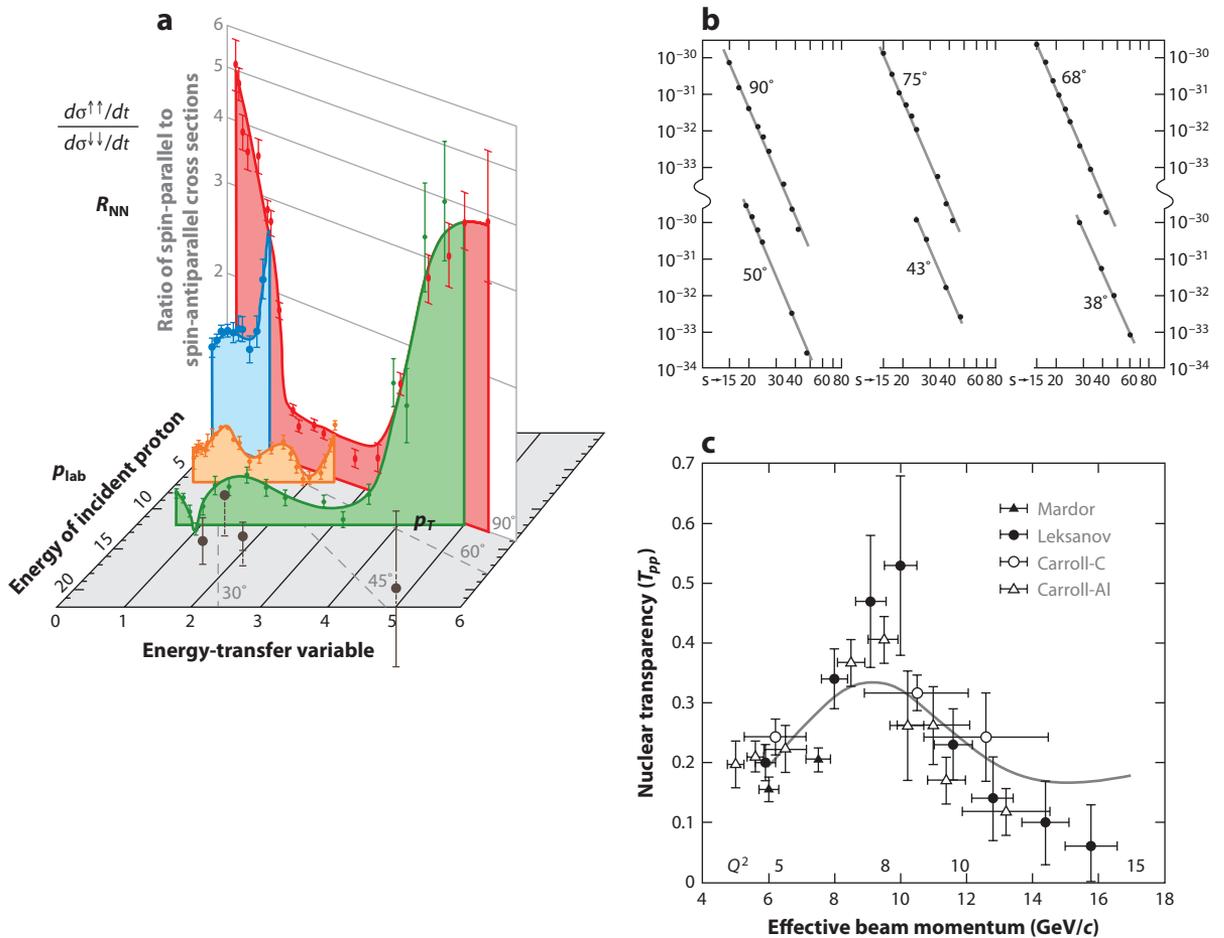

**Figure 2**

(*a*) Measurement of $R_{NN}$, the ratio of transverse spin-parallel to transverse spin-antiparallel elastic *pp* scattering (14) as a function of momentum transfer. (*b*) The $s^{-10}$ scaling of the measured elastic $pp \to pp$ scattering cross section at a fixed center-of-mass angle. (*c*) Transparency ratio in quasi-elastic *pp* scattering in nuclei (20). References to the experiments and theory fit can be found in Reference 20. Panel *a* reproduced with permission. Copyright 2012, Scientific American, Inc. All rights reserved.

protons are polarized normal to the scattering plane. Remarkably, the ratio of spin-parallel to spin-antiparallel scattering, $R_{NN}$, reaches 4:1 at $\sqrt{s} \simeq 5$ GeV (**Figure 2a**).

Proton-proton elastic scattering is a well-measured process. The unpolarized differential cross section, $\frac{d\sigma}{dt}(pp \to pp)$, follows the pQCD prediction (15, 16), $s^{10}\frac{d\sigma}{dt} \simeq F(\theta_{cm})$, over the entire domain of hard scattering accessed by experiments (**Figure 2b**). However, a tour de force measurement by Krisch and colleagues (14) finds a unexpectedly large spin-spin correlation at $p_{lab} = 12.7$ GeV/*c*; in other words, $\sqrt{s} \simeq 5$ GeV. Remarkably, the cross section when both protons are polarized parallel and normal to the scattering plane rises rapidly to more than four times the cross section when the proton spins are antiparallel (**Figure 2a**). This is the largest spin-spin correlation ever observed in hadron physics; it strongly contradicts pQCD expectations. Because the natural scale $\Lambda_{\overline{MS}}$ for QCD is only a few hundred MeV, it is hard to understand why deviations from pQCD should have an onset at such a large mass scale: $p_T^2 = (tu/s) \simeq 6$ GeV$^2$.

One can measure large-angle quasi-elastic *pp* scattering on the bound protons in a nucleus. In conventional Glauber theory, only the $Z^{1/3}$ nucleons on the periphery of the target nucleus are predicted to interact because the absorptive cross section is so large. In contrast, in pQCD the nucleons, which scatter at high $p_T$, interact when their wave function has a small transverse size: $b_\perp \sim (1/p_T)$. These fluctuations have a small color-dipole moment that allows interaction with all the $Z$ protons in the target nucleus. pQCD thus predicts that the hard-scattered hadron suffers minimal absorption as it transits the nucleus (17). This remarkable phenomenon has been observed and confirmed quantitatively in reactions such as diffractive dijet production, $\pi A \to$ jet, jet, $A'$ (18), and quasi-elastic electroproduction of vector mesons (19). Color transparency is a clear manifestation of the role of gauge interactions in hadron physics. In fact, pQCD color transparency has also been clearly observed (20) in quasi-elastic proton scattering, in which the effective number of interacting protons in the target increases with $p_T$. However, this observation is true only at energies below $\sqrt{s} \sim 5$ GeV; in fact, color transparency disappears at the same energy and angles that show the anomalous spin-spin correlation (**Figure 2c**).

What could cause the simultaneous appearance of the large spin-spin correlation and the breakdown of color transparency in large-angle $pp \to pp$ scattering at $\sqrt{s} \simeq 5$ GeV? Note that this is also the energy for producing hidden charm at threshold in the intermediate state—for example, the formation of an octoquark ($|uuduudc\bar{c}\rangle$) resonance (**Figure 3b**) (21). The natural quantum number in the *pp* amplitude for the lowest-mass resonance is $J = 1 = L = S$ with negative parity, given that the $c$ and the $\bar{c}$ have opposite parity. Remarkably, the protons can form this state only if $R_{NN} = \infty$. The interference of the resonance amplitude with the background quark-interchange amplitude provides a reasonable fit to the kinematic behavior of the $pp \to pp$ rate at large angles (**Figure 3a**). The production cross section for charm at threshold in *pp* collisions is predicted to be 1μb, which is compatible with unitarity and analyticity (22). Enhanced charm production is also predicted at threshold in reactions such as $\gamma p \to J/\psi p$ and $\gamma p \to D\bar{D}p$, which will be accessible at the new 12-GeV JLab. A comparable effect is also observed at the $\phi$ threshold (21).

### 3.1. Nuclear-Bound Quarkonium

The existence of the $|uuduudc\bar{c}\rangle$ octoquark state illustrated in **Figure 3b** may resemble a $J/\psi$ *pp* system and, thus, would be the first example of another novel QCD phenomenon: a nuclear-bound quarkonium (23). The QCD analog of an atomic molecule is a bound state of heavy quarkonium with a nucleus, such as $[J/\psi A]$ (23, 24). The binding occurs through two-gluon exchange, the hadronic analog of the van der Waals interaction. Because both the kinetic energy of the $J/\psi$ and that of the nucleus are small, one would expect these exotic hybrid states to be produced at threshold. The binding of charmonium in nuclear matter is approximately 10 MeV, which is comparable with the binding energy of nucleons in nuclei—a remarkable result, given that the nucleon-nucleon interaction is two orders of magnitude stronger than the $c\bar{c} - N$ interaction (22). This difference arises from the absence of Pauli blocking in the charmonium-nucleon system (23). Examples of nuclear-bound quarkonium are the $|uuduuds\bar{s}\rangle$ and $|uuduudc\bar{c}\rangle$ resonances, which apparently contribute as intermediate states in $pp \to pp$ elastic exchange. These resonances can account (21) for the large spin-spin correlations (14) observed at both the strangeness ($E_{cm} \simeq 3$ GeV) and charm ($E_{cm} \simeq 5$ GeV) thresholds (**Figure 3a**).

### 4. DOMINANCE OF QUARK INTERCHANGE

Hadron scattering amplitudes at high-momentum transfer factorize in pQCD into a hard-scattering amplitude $T_H$, which describes the scattering and rearrangement of the constituent

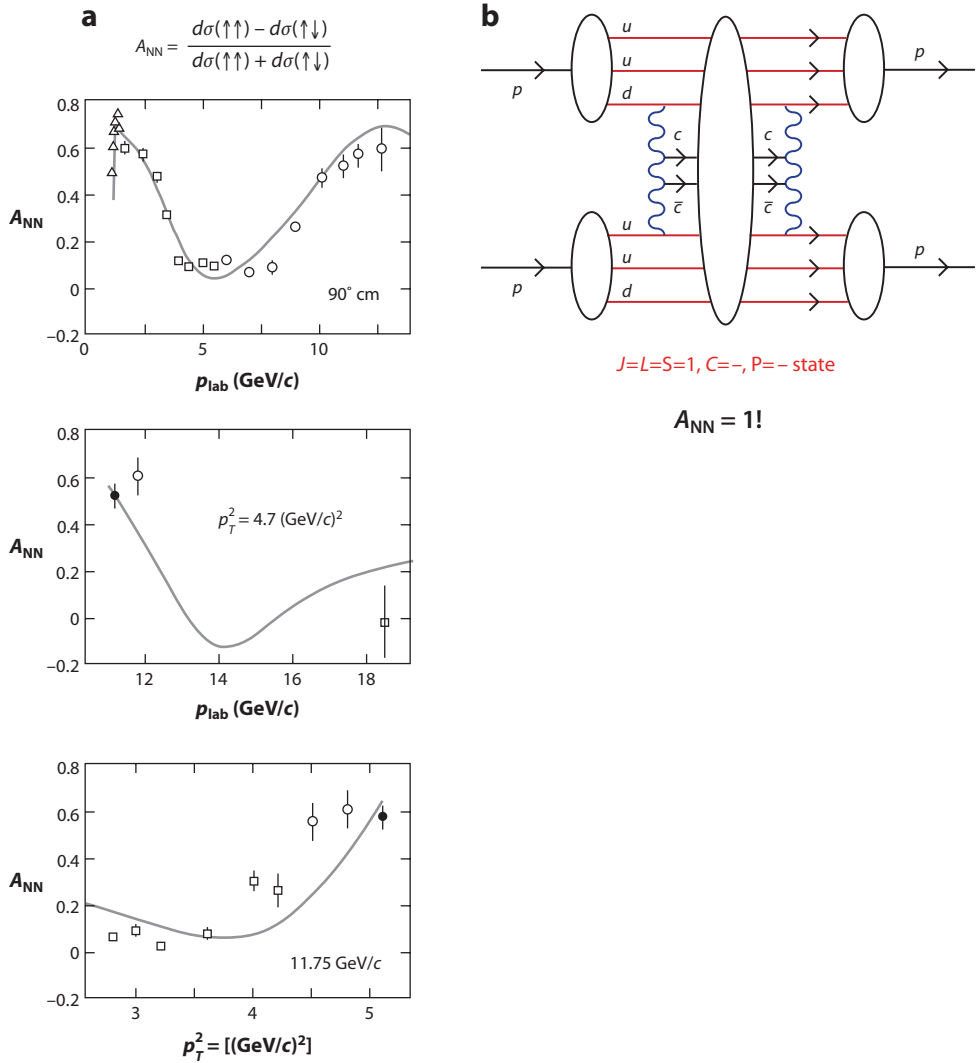

**Figure 3**

(*a*) Predictions (22) of the octoquark threshold model for the transverse spin-parallel to transverse spin-antiparallel asymmetry $A_{NN} = \frac{d\sigma_{\uparrow\uparrow} - d\sigma_{\uparrow\downarrow}}{d\sigma_{\uparrow\uparrow} + d\sigma_{\uparrow\downarrow}}$ in polarized elastic *pp* scattering. The model includes $J^P = 1^-$ octoquark resonances at both the $M = 3$ GeV strangeness and $M = 5$ GeV charm thresholds, together with perturbative quantum chromodynamics quark-interchange amplitude. (*b*) The $J = 1$, $L = 1$, $S = 1$ octoquark resonance at the $c\bar{c}$ threshold in elastic *pp* scattering.

quarks convoluted with hadron distribution amplitudes, $\phi(x_i, Q)$, for each hadron (16). One would normally expect gluon exchange diagrams to dominate large-angle elastic-scattering exclusive hadron-hadron scattering reactions. For example, as shown by Landshoff (25), gluon exchange implies that large-angle *pp* elastic scattering is dominated by a sequence of three $qq \to qq$ amplitudes, each of which has a small gluon virtuality, $t/9$. The Landshoff mechanism predicts $\frac{d\sigma}{dt}(pp \to pp) \propto (1/t^8)$; in fact, measurements are consistent with $\frac{d\sigma}{dt}(pp \to pp) \propto (1/s^2 u^4 t^4)$, as predicted by the quark-interchange mechanism (26, 27).

White et al. (28) showed that two-body scattering amplitudes at fixed $\theta_{\rm cm}$ are dominated by the quark exchange and interchange amplitudes (26), rather than by gluon exchange contributions. The quark interchange amplitude is the analog of spin exchange in atom-atom scattering, wherein the scattering occurs via the exchange of a common constituent. For example, $K^+ p \to K^+ p$ elastic scattering can occur by the interchange of the valence $u$ quark in the kaon with one of the valence $u$ quarks in the proton. The interchange amplitude can be written in terms of an overlap of hadronic light-front (LF) wave functions (26):

$$M(s,t)_{(AB \to CD)} = \int d^2 k_\perp \int_0^1 \frac{dx}{16\pi^3 x^2 (1-x)^2} \Delta \quad\quad 3.$$
$$\psi_A(x, \vec{k}_\perp - x\vec{r}_\perp + (1-x)\vec{q}_\perp)\psi_C(\vec{k}_\perp - x\vec{r}_\perp)\psi_B(x, \vec{k}_\perp)\psi_D(\vec{k}_\perp + (1-x)\vec{q}_\perp),$$

where $u = \vec{r}_\perp^2$, $t = \vec{q}_\perp^2$, $\vec{r}_\perp \cdot \vec{q}_\perp = 0$, and $s + t + u = M_A^2 + M_B^2 + M_C^2 + M_D^2$. Here, $x$ is the LF fraction of the exchanged quark. The quantity

$$\Delta = s - M_a^2 - M_b^2 - K_a - K_b - K_c - K_d \quad\quad 4.$$

is the analog of a potential $V = E - T$ with LF kinetic energies, $K_q = (k_{\perp q}^2 + m_q^2)/x$, for each constituent quark $q$ in the reaction. The resulting quark interchange prediction for the $K^+ p \to K^+ p$ amplitude at large momentum transfer is $1/ut^2$, so

$$\frac{d\sigma}{dt}(K^+ p \to K^+ p) \propto \frac{1}{s^2 u^2 t^4}, \quad\quad 5.$$

which agrees with the angular distribution and fixed-angle $s^{-8}$ scaling of the measured differential cross section.

Thus, one has a hadron physics puzzle: Why is gluon exchange absent in every measured hard-scattering exclusive hadron reaction? Note that in the Landshoff process, the virtualities of the exchanged gluons in the measured processes are typically less than 1 GeV$^2$. The reason could be the effective absence of soft gluon quanta at the hadronic scale (7). This result is consistent with the flux-tube interpretation of QCD (6), in which soft gluons interact so strongly that they are sublimated into a color-confinement potential for quarks. A similar scenario also appears in the AdS/QCD holographic model for hadron physics (Section 18). In this model, higher Fock states can have any number of extra $q\bar{q}$ pairs but, surprisingly, no dynamical gluons. This unusual property of AdS/QCD may therefore explain the dominance of quark interchange (26) in large-angle elastic scattering.

## 5. THE UNEXPECTED ROLE OF DIRECT PROCESSES IN HIGH-$p_T$ HADRON REACTIONS

It is normally assumed that hadrons produced at high transverse momentum in inclusive high-energy hadronic collisions such as $pp \to HX$ arise only from jet fragmentation. A fundamental test of leading-twist QCD predictions in high–transverse momentum hadronic reactions is the measurement of the power-law falloff of the inclusive cross section (29): $E d\sigma/d^3 p(AB \to CX) = F(\theta_{\rm cm}, x_T)/p_T^{n_{\rm eff}}$ at fixed $x_T = 2p_T/\sqrt{s}$ and fixed $\theta_{\rm cm}$. In the case of the scale-invariant parton model, $n_{\rm eff} = 4$. However, in QCD $n_{\rm eff} \sim 4 + \delta$, where $\delta \simeq 1.5$ is the typical correction to the conformal prediction arising from the QCD running coupling and the DGLAP (Gribov–Lipatov–Dokshitzer–Altarelli–Parisi) evolution of the input parton distribution and fragmentation functions (30, 31).

The usual expectation is that leading-twist subprocesses (i.e., the leading power-law contributions) dominate measurements of high-$p_T$ hadron production at RHIC and Tevatron energies. In

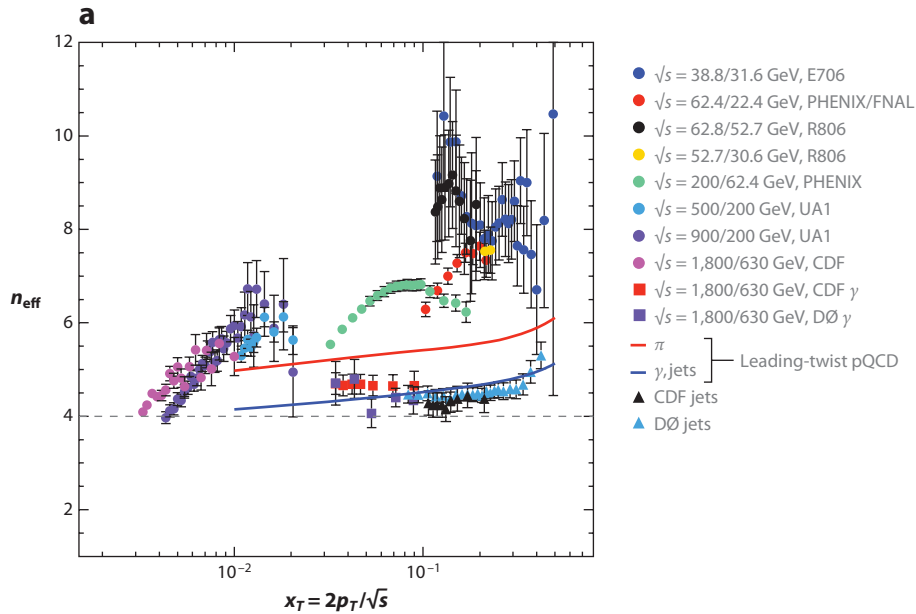

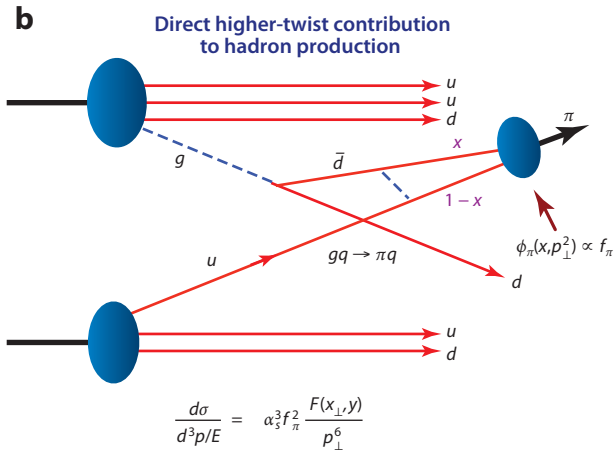

**Figure 4**

(*a*) Scaling of the inclusive cross sections for hadrons, photons, and jets at high $p_T$ at fixed $x_T = 2\frac{p_T}{\sqrt{s}}$. This comparison between experiment and leading-twist perturbative QCD (pQCD) is given in References 30 and 31. (*b*) Example of a direct QCD contribution for pion production.

fact, the data obtained from isolated photon production $pp \to \gamma_{\text{direct}} X$, as well as for jet production, agree well with the leading-twist scaling prediction, $n_{\text{eff}} \simeq 4.5$ (30). However, measurements of $n_{\text{eff}}$ for hadron production are not consistent with the leading-twist predictions (**Figure 4***a*). Striking deviations from the leading-twist predictions have also been observed at lower energy at the ISR (Intersecting Storage Rings) and Fermilab fixed-target experiments (29, 32, 33). Such deviations point to a significant contribution from direct higher-twist processes in which the hadron is created directly in the hard subprocess, rather than from quark or gluon jet fragmentation.

A significant fraction of high-$p_\perp^H$ isolated hadrons can emerge directly from hard higher-twist subprocesses (30, 31), even at the LHC. An example is shown in **Figure 4b**. The direct production of hadrons can also explain (34) the remarkable baryon anomaly observed at RHIC: Both the ratio of baryons to mesons at high $p_\perp^H$ and the power-law falloff $1/p_\perp^n$ at fixed $x_\perp = 2p_\perp/\sqrt{s}$ increase with centrality (35), in contrast to the usual expectation that protons should suffer more energy loss in the nuclear medium than mesons do. The high values of $n_{\text{eff}}$ with $x_T$ observed in the data indicate the presence of an array of higher-twist processes, including subprocesses in which the hadron enters directly, rather than through jet fragmentation (36). Although they are suppressed by powers of $1/p_T$, the direct higher-twist processes can dominate because they are energy efficient—no same-side energy or momentum is lost from the undetected fragments. Thus, the incident colliding partons are evaluated at the minimum possible values of LF momentum fractions $x_1$ and $x_2$, wherein the parton distribution functions are numerically large.

### 5.1. Direct Subprocesses and the Drell–Yan Reaction

Direct processes in which a hadron appears in the hard-scattering subprocess (37) can explain why angular distribution of the lepton pair in the Drell–Yan process $\pi p \to \ell^+\ell^- X$ changes from the conventional $1 + \cos^2\theta$ prediction, characteristic of the $q\bar{q} \to \ell^+\ell^-$ process, to $\sin^2\theta$ as the longitudinal-momentum fraction of the lepton pair approaches $x_F \to 1$. In fact, at high $x_F$, the momenta of both the valence quark and the antiquark are required. Thus, one must include direct subprocesses such as $\pi q \to \ell^+\ell^- q$. The resulting angular distribution of the lepton in the lepton pair center of mass is $\sin^2\theta$, which reflects the spin of the pion. Because the distribution amplitude of the pion enters, this higher-twist process is suppressed by a factor of $f_\pi^2/Q^2$; nevertheless, because it does not fall off at high $x_F$, it dominates over the conventional $(1 - x_F)^2$ contribution in the high-$x_F$ domain.

### 5.2. The RHIC Baryon Anomaly

Normally, many more pions than protons are produced at high transverse momentum in hadron-hadron collisions. This is also true for peripheral collisions of heavy ions. However, when the nuclei collide with maximal overlap (i.e., in central collisions) the situation is reversed: More protons than pions emerge. This observation at RHIC (35) contradicts the usual expectation that protons should be more strongly absorbed than pions in the nuclear medium. This deviation also points to a significant contribution from direct higher-twist processes in which hadrons, particularly baryons, are created directly in the hard subprocess rather than from quark or gluon jet fragmentation. Because these processes create color-transparent baryons, this mechanism can explain the RHIC baryon anomaly (34).

### 5.3. Anomalous $J/\psi$ Polarization at High Transverse Momentum

The standard pQCD prediction is that quarkonium states at high $p_T$ are produced dominantly from fragmentation of gluon jets. Because the gluon has transverse polarization, one would also predict transverse polarization of the $J/\psi$ and other quarkonium states (38, 39). This expectation is contradicted by experimental results, which indicate that other QCD production mechanisms are important (40).

## 6. INTRINSIC HEAVY QUARKS

It is conventional to assume that the charm and bottom quarks in the proton structure functions arise only from gluon splitting $g \to Q\bar{Q}$. For example, the proton wave function at soft scales is assumed to contain only the valence quarks and gluons. DGLAP evolution from the $g \to Q\bar{Q}$ splitting process then generates all of the sea quarks at virtuality $Q^2 > 4m_Q^2$. If this hypothesis were correct, then the $\bar{u}(x)$ and $\bar{d}(x)$ distributions would be identical. Similarly, if sea quarks arise only from gluon splitting, one would expect the $s(x)$ and $\bar{s}(x)$ distributions to be the same. However, measurements of Drell–Yan processes, deep-inelastic electron and neutrino scattering, and other experiments show that these predictions are false. This also leads to a violation of the Gottfried sum rule (41).

In the LF Fock state approach, sea quarks are identified with five-quark $|uudq\bar{q}\rangle$ Fock state wave functions in which strong QCD interactions appear, particularly at equal rapidity. The proton LF wave function contains ab initio intrinsic heavy quark Fock state components such as $|uudc\bar{c}\rangle$ (**Figure 5b**) (42–45). The intrinsic heavy quarks carry most of the proton's momentum because doing so minimizes the off-shellness of the state. The heavy quark pair $Q\bar{Q}$ in the intrinsic Fock state is primarily a color octet, and the ratio of intrinsic charm to intrinsic bottom scales as $m_c^2/m_b^2 \simeq 1/10$, as readily observed from the operator product expansion in non-Abelian QCD (43, 45). Intrinsic charm and bottom explain the origin of high-$x_F$ open-charm and open-bottom hadron production, as well as the single and double $J/\psi$ hadroproduction cross sections observed at high $x_F$ and the factorization-breaking nuclear $A^\alpha(x_F)$ dependence of hadronic $J/\psi$ production cross sections.

Intrinsic heavy quarks provide a novel mechanism for the inclusive and diffractive Higgs production $pp \to ppH$, in which the Higgs boson carries a significant fraction of the projectile proton momentum (46, 47). The production mechanism is based on the subprocess $(Q\bar{Q})g \to H$, in which the Higgs boson acquires the momentum of the $Q\bar{Q}$ pair in the $|uudQ\bar{Q}\rangle$ intrinsic heavy quark Fock state of the colliding proton and thus has approximately 80% of the projectile proton's momentum. The high-$x_F$ Higgs boson could be accessed at the LHC by far-forward detectors or by arranging the proton beams to collide at a significant crossing angle.

The analog of intrinsic charm in hadrons is the $\mu^+\mu^-$ content of positronium. The $|e^+e^-\mu^+\mu^-\rangle$ Fock state appears through the cut of the muon-loop light-by-light contribution to the self-energy of the positronium eigenstate. In this Fock state, the muons carry almost all of the momentum of the moving atom because the off-shell virtuality is minimal at equal velocity. In QED, the probability that intrinsic leptons $L\bar{L}$ exist in positronium scales as $1/m_L^4$, whereas in QCD, the probability of intrinsic heavy quarks in the wave function of a light hadron scales as $1/m_Q^2$ because of its non-Abelian couplings (43, 45).

The five-quark Fock state of the proton's LF wave function $|uudQ\bar{Q}\rangle$ is thus the primary origin of the sea quark distributions of the proton (48, 49). Experiments show that the sea quarks have remarkable nonperturbative features, such as $\bar{u}(x) \neq \bar{d}(x)$, and an intrinsic strangeness (50) distribution, $s(x)$, appearing at $x > 0.1$, as well as intrinsic charm and bottom distributions at high $x$. Such distributions (43, 45) arise rigorously from $gg \to Q\bar{Q} \to gg$ insertions connected to the valence quarks in the proton self-energy; in fact, they fit a universal intrinsic quark model (**Figure 5a**) (42), as shown by Chang & Peng (51).

### 6.1. The DØ Measurement of $\bar{p}p \to c + \gamma + X$

The DØ Collaboration (52) at the Tevatron has measured the processes $\bar{p}p \to c + \gamma + X$ and $\bar{p}p \to b + \gamma + X$ at very high photon transverse momentum: $p_T^\gamma \sim 120\,\text{GeV}/c$. The rate for

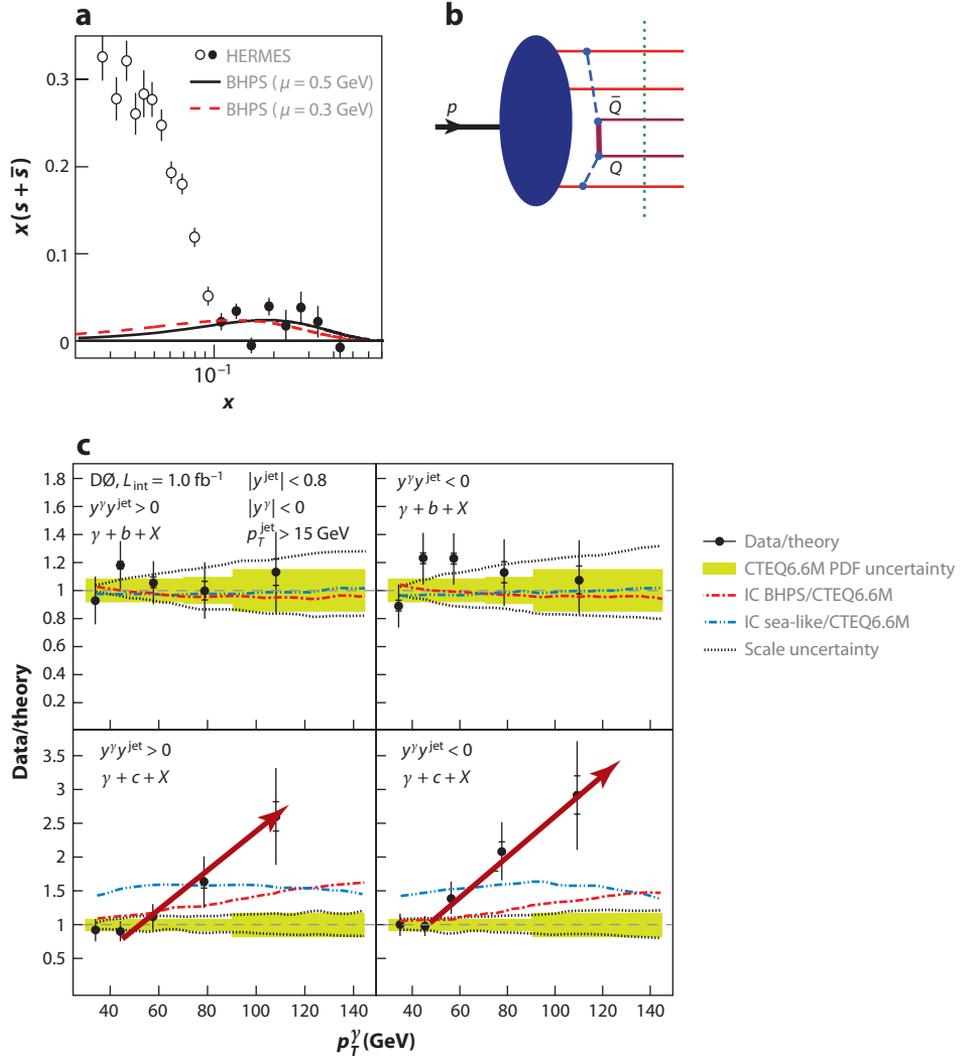

**Figure 5**

(*a*) Intrinsic and extrinsic strangeness distribution (51), obtained by use of the model from Reference 42. (*b*) Five-quark Fock state of the proton and origin of the intrinsic sea. (*c*) DØ measurement of proton-antiproton annihilation at the Tevatron into a high–transverse momentum photon plus a bottom or charm quark jet. The two hemispheres are plotted separately. A comparison between two intrinsic charm (IC) parton distribution functions (PDFs), assuming massless charm quark evolution (53), is also shown.

$\bar{p}p \to b + \gamma X$ agrees very well with next-to-leading-order (NLO) pQCD predictions (**Figure 5***c*); however, the corresponding charm cross section strongly deviates from the standard pQCD prediction given above: $p_T^\gamma \sim 60$ GeV/$c$. This anomaly requires an intrinsic contribution to the charm structure function at $Q^2 \sim 10^4$ GeV$^2$—a factor of two larger than the standard parameterizations (42, 53). The reason for this discrepancy could be that DGLAP evolution of the intrinsic charm component is significantly reduced because of the suppression of the $c \to cg$ radiative process for heavy quarks.

## 6.2. SELEX Double-Charm Isospin Problem

The SELEX Collaboration (54) has reported a discovery of a set of doubly charmed spin-1/2 and spin-3/2 baryons with quantum numbers matching |ccu⟩ and |ccd⟩ bound states. However, the measured mass splittings of the *ccu* and *ccd* states are much larger than expected from known QCD isospin-splitting mechanisms. A speculative proposal (55) is that these baryons have a linear configuration *cqc* in which the light quark *q* is exchanged between the heavy quarks as in a linear molecule. The linear configuration enhances the Coulomb repulsion of the *cuc* relative to *cdc*. Clearly, it is important to obtain experimental confirmation of the SELEX results.

## 6.3. The Anomalous Factorization-Breaking Nuclear Dependence of $J/\psi$ Hadroproduction

The cross section for $J/\psi$ production in a nuclear target has been well measured. The ratio of the nuclear and proton target cross sections has the form $A^{\alpha(x_F)}$, where $x_F$ is the Feynman fractional longitudinal momentum of the $J/\psi$. At low $x_F$, $\alpha(x_F)$ is slightly smaller than one, but at $x_F \sim 1$, it decreases to $\alpha = 2/3$. These results (**Figure 6a**) are surprising because (*a*) the value $\alpha = 2/3$ is characteristic of a strongly interacting hadron, not a small-size quarkonium state, and (*b*) the functional dependence $A^{\alpha(x_F)}$ contradicts pQCD factorization predictions.

This anomaly, in combination with the anomalously large and flat cross sections measured at high $x_F$, is consistent with a QCD mechanism based on color-octet intrinsic charm Fock states: Because of its large color dipole moment, the intrinsic heavy quark Fock state of the proton, $|(uud)_{8_C}(c\bar{c})_{8_C}\rangle$, interacts primarily with the $A^{2/3}$ nucleons at the front surface (**Figure 6b**). The $c\bar{c}$ color octet thus scatters on a front-surface nucleon, changes to a color singlet, and then propagates through the nucleus as a $J/\psi$ at high $x_F$.

## 7. THE ROLE OF QUANTUM CHROMODYNAMICS HIDDEN COLOR IN NUCLEAR REACTIONS

In nuclear physics, nuclei are composites of nucleons. However, QCD provides a new perspective (56, 57): Six quarks in the fundamental $3_C$ representation of $SU(3)$ color can combine into five different color-singlet combinations, only one of which corresponds to a proton and a neutron. The deuteron wave function is a proton-neutron bound state at large distances, but as the quark separation decreases, QCD evolution due to gluon exchange introduces four other, hidden-color states into the deuteron wave function (58). The normalization of the deuteron form factor observed at high $Q^2$ (59), as well as the presence of two mass scales in the scaling behavior of the reduced deuteron form factor (56), suggests sizable hidden-color Fock state contributions in the deuteron wave function (60) that are much larger than expected from the small, 2.2-MeV binding energy of the deuteron (**Figure 7**). The hidden-color states of the deuteron have significant internal energy and can be materialized at the hadron level as $\Delta^{++}(uuu)$, $\Delta^{-}(ddd)$, and other novel quantum fluctuations of the deuteron. These massive dual hadronic components become important as one probes the deuteron at short distances, such as in exclusive reactions at large momentum transfer. For example, the ratio $d\sigma/dt(\gamma d \to \Delta^{++}\Delta^{-})/d\sigma/dt(\gamma d \to np)$ is predicted to increase to a fixed ratio of 2:5 with increasing transverse momentum $p_T$. Similarly, the Coulomb dissociation of the deuteron into various exclusive channels $ed \to e' + pn, pp\pi^-, \Delta\Delta, \ldots$ is expected to change in composition as the final-state hadrons are probed at high transverse momentum, reflecting the onset of hidden-color degrees of freedom.

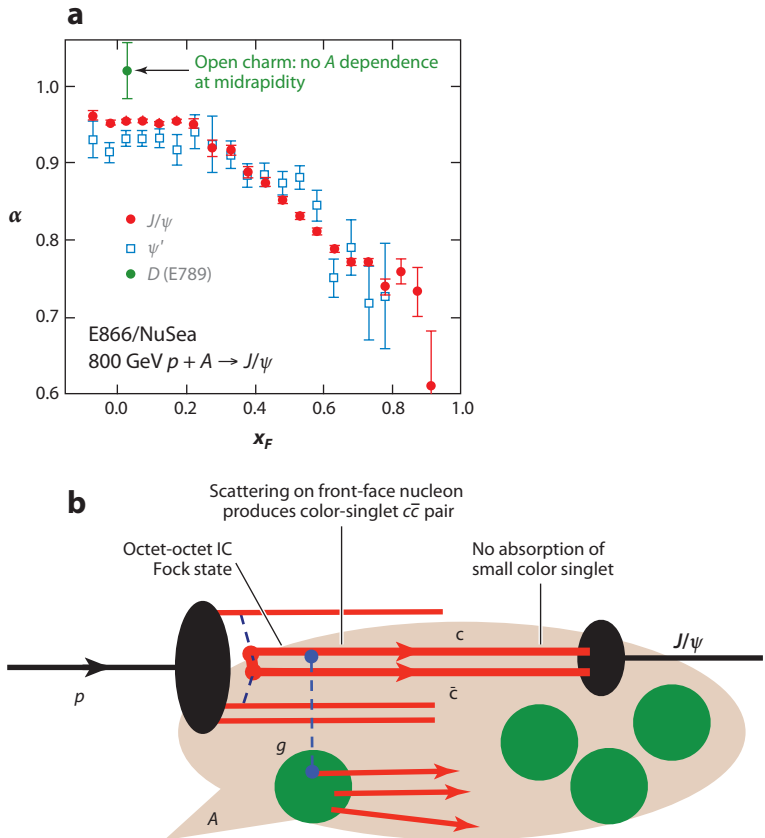

**Figure 6**

(*a*) E866/NuSea data for the nuclear *A* dependence of open and hidden charm hadroproduction as a function of the Feynman longitudinal momentum fraction $x_F$. (*b*) Intrinsic charm (IC) model (46, 47) for the *A* dependence of *J*/$\psi$ hadroproduction.

The hidden color of the deuteron can be probed in electron-deuteron collisions through the study of reactions such as $\gamma^* d \to npX$, in which the proton and neutron emerge in the target fragmentation region at high and opposite $p_T$. In principle, one can also study deep-inelastic lepton scattering (DIS) reactions $ed \to e'X$ at very high $Q^2$, where $x > 1$. The production of high-$p_T$ antinuclei such as $pp \to \bar{d}X$ is also sensitive to hidden-color nuclear components.

## 8. BREAKDOWN OF PERTURBATIVE QUANTUM CHROMODYNAMICS FACTORIZATION THEOREMS

The factorization scenario from the parton model and then from pQCD has played a guiding role in virtually all aspects of hadron physics phenomenology. In the case of inclusive reactions such as $E_H d\sigma/d^3 p_H)(pp \to HX)$, the pQCD ansatz predicts that the cross section at leading order (LO) in the transverse momentum $p_T$ can be computed by convoluting the perturbatively calculable hard subprocess quark and gluon cross section with the process-independent structure functions of the colliding hadrons and the quark fragmentation functions. The resulting cross section scales as $1/p_T^4$, modulo the DGLAP scaling violations derived from the logarithmic evolution of the

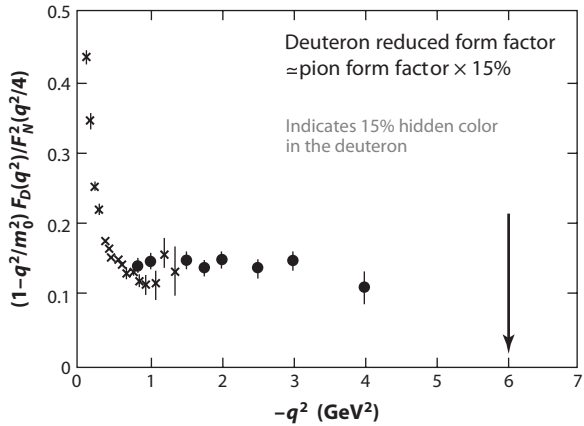

**Figure 7**

The ratio of the reduced deuteron form factor to the pion form factor (56, 59), indicating a significant fraction of hidden color (56–58) in the deuteron wave function.

structure functions and fragmentation distributions, as well as the running of the QCD coupling appearing in the hard-scattering subprocess matrix element.

The effects of the final-state interactions of the scattered quark in deep-inelastic scattering have traditionally been assumed to give either an inconsequential phase factor or power law–suppressed corrections. However, this is true only for sufficiently inclusive cross sections. For example, consider semi-inclusive DIS (SIDIS) on a polarized target, $\ell p_\updownarrow \to H \ell' X$. In this case, the final-state gluonic interactions of the scattered quark lead to a $T$-odd nonzero spin correlation of the plane of the lepton-quark scattering plane with the polarization of the target proton (61), which is not power-law suppressed with increasing virtuality of the photon $Q^2$; that is, it exhibits Bjorken scaling. This leading-twist Sivers effect (62) is nonuniversal in that pQCD predicts an opposite-sign correlation in Drell–Yan reactions relative to single-inclusive deep-inelastic scattering (63, 64). This important but as-yet-untested prediction arises because the Sivers effect in the Drell–Yan reaction is modified by the initial-state interactions of the annihilating antiquark.

Similarly, the final-state interactions of the produced quark with its comoving spectators in SIDIS produce a final-state $T$-odd polarization correlation: the Collins effect. This effect can be measured without beam polarization by measuring the correlation of the polarization of a hadron such as the $\Lambda$ baryon with the quark-jet production plane. Analogous spin effects occur in QED reactions due to rescattering via final-state Coulomb interactions. Although the Coulomb phase for a given partial wave is infinite, the interference between Coulomb phases arising from different partial waves leads to observable effects. These considerations have led to a reappraisal of the range of validity of the standard factorization ansatz (65).

**Figure 8** illustrates the calculation of the Sivers single-spin asymmetry in DIS in QCD. The analysis requires two different orbital angular-momentum components: an $S$-wave with the quark-spin parallel to the proton spin and a $P$-wave for the quark with antiparallel spin. The difference between the final-state Coulomb phases leads to a $\vec{S} \times \vec{q} \times \vec{p}$ correlation of the proton's spin with the virtual photon-to-quark production plane (61). Thus, as made clear by its QED analog, the final-state gluonic interactions of the scattered quark lead to a $T$-odd nonzero spin correlation of the plane of the lepton-quark scattering plane with the polarization of the target proton (61).

The $S$- and $P$-wave proton wave functions also appear in the calculation of the Pauli form factor quark by quark. Thus, one can correlate the Sivers asymmetry for each struck quark with the

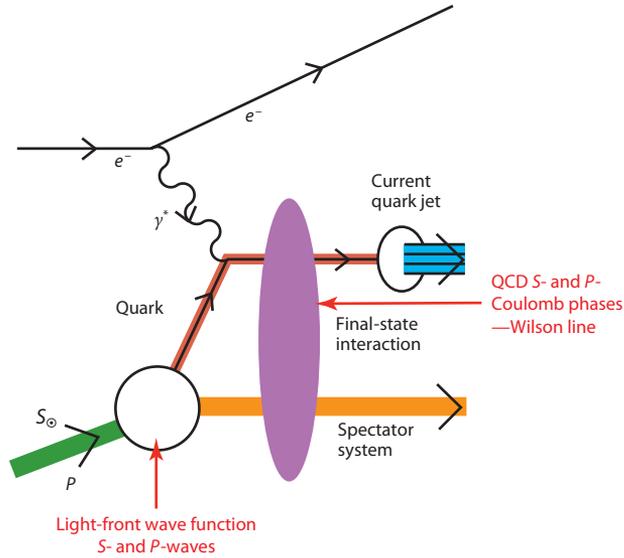

**Figure 8**

Origin of the leading-twist Sivers single-spin asymmetry in deep-inelastic lepton scattering in quantum chromodynamics (QCD). The lensing effect of the final-state interactions in two different partial waves leads to a pseudo-$T$-odd correlation between the transverse proton spin and the production plane. The analogous initial-state correction leads to the same correlation but with an opposite sign in Drell–Yan reactions (64, 65).

anomalous magnetic moment of the proton carried by that quark (66), which leads to the prediction that the Sivers effect is larger for positive pions, as observed by the HERMES experiment at DESY (8), the COMPASS experiment (9–11) at CERN, and the CLAS experiment at JLab (12, 13).

The physics of the lensing dynamics or Wilson-line physics (67) underlying the Sivers effect involves nonperturbative quark-quark interactions at low momentum transfer, not the hard scale $Q^2$ of the virtuality of the photon. It would be interesting to learn whether the strength of the soft initial- or final-state scattering can be predicted by using the effective confining potential of QCD from LF holographic QCD described in Section 18.

Measurements (68) of the Drell–Yan process $\pi p \to \mu^+\mu^- X$ reveal an angular distribution that contradicts pQCD expectations. In particular, there is an anomalously large $\cos 2\phi$ azimuthal angular correlation between the lepton decay plane and its production plane, which contradicts the Lam–Tung relation, a prediction of pQCD factorization (69). Again such effects point to the importance of initial- and final-state interactions of the hard-scattering constituents (70), corrections that are not included in the standard pQCD factorization formalism.

As noted by Collins & Qiu (65), the details of the traditional factorization formalism of pQCD fail for many hard inclusive reactions because of initial- and final-state interactions. For example, if both the quark and the antiquark in the Drell–Yan subprocess $q\bar{q} \to \mu^+\mu^-$ interact with the spectators of the other hadron, then there should be a $\cos 2\phi \sin^2\theta$ planar correlation in unpolarized Drell–Yan reactions (70). This double Boer–Mulders effect can account for the anomalously large $\cos 2\phi$ correlation and the corresponding violation (70, 71) of the Lam–Tung relation for Drell–Yan processes observed by the NA10 Collaboration (68). These are additional signals of initial- and final-state interactions. One also observes large single-spin asymmetries in reactions such as $pp \leftrightarrow \pi X$, an effect that has not yet been explained (72). Another important signal for factorization breakdown at the LHC will be the observation of a $\cos 2\phi$ planar correlation in dijet production.

The final-state interactions of the struck quark with the spectators (73) also lead to diffractive events in deep-inelastic scattering at leading twist such as $\ell p \to \ell' p' X$, in which the proton remains intact and isolated in rapidity; in fact, approximately 10% of the deep-inelastic lepton-proton scattering events observed at HERA are diffractive (74, 75). This finding is surprising given that the underlying hard subprocess $\ell q \to \ell' q'$ greatly disrupts the target nucleon. The existence of a rapidity gap between the target and the diffractive system requires that the target remnant emerge in a color-singlet state, which is made possible in any gauge by the soft rescattering incorporated in the Wilson line or by augmented LF wave functions. Quite different fractions of single-diffractive $pp \to$ jet $p'X$ events and double-diffractive $p\bar{p} \to$ jet $p'\bar{p}'X$ events have been observed at the Tevatron. The underlying mechanism is believed to be soft gluon exchange between the scattered quark and the remnant system in the final state, which occurs following hard scattering.

By using Gribov–Glauber theory, one can show (76) that the Bjorken-scaling diffractive deep-inelastic scattering events lead to the shadowing of nuclear structure functions at low $x_{Bjorken}$. This shadowing is due to the destructive interference of two-step and one-step amplitudes in the nucleus. Because diffraction involves rescattering, shadowing and diffractive processes are not intrinsic properties of hadron and nuclear wave functions and structure functions; rather, they are properties of the complete dynamics of the scattering reaction (77).

### 8.1. The $t\bar{t}$ Asymmetry Observed at the Tevatron

The CDF (78) and DØ (79) Collaborations at the Tevatron have recently reported that the $t$ and $\bar{t}$ heavy quarks do not have the same momentum distributions in $\bar{p}p \to t\bar{t}X$ events. The observed asymmetry is much larger than predicted from QCD NLO corrections to the $\bar{q}q \to t\bar{t}$ subprocess. This discrepancy may indicate new physics such as the existence of so-called axigluons, new heavy particles that couple to quarks with positive charge conjugation. Alternatively, the Tevatron $t\bar{t}$ asymmetry may indicate the importance of rescattering Coulomb-like final-state interactions of the top quarks with $ud$ and $\bar{u}\bar{d}$ remnant spectators of the colliding proton and antiproton (**Figure 9**) (S.J. Brodsky, B. von Harling & Y. Zhao, manuscript in preparation). This effect could also lead to a $t\bar{t}$ asymmetry in $pp \to t\bar{t}X$ collisions at the LHC because the $t$ quark may be color attracted to one of the spectator *ud* diquarks produced in the $q\bar{q} \to t\bar{t}$ subprocess; however, the effect would be significant only when the *t* and *ud* systems have a small rapidity separation.

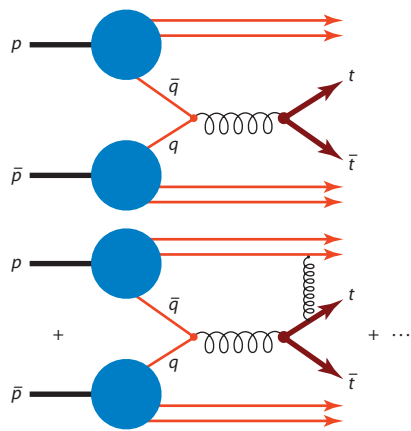

**Figure 9**
The interference between Born and rescattering amplitudes provides a top quark/anti–top quark asymmetry in hadron-hadron collisions.

## 9. NONUNIVERSAL ANTISHADOWING

It is usually assumed that the nuclear modifications to the structure functions measured in deep-inelastic charged lepton–nucleus and neutrino-nucleus interactions are identical. In fact, Gribov–Glauber theory predicts that the antishadowing of nuclear structure functions is not universal but rather depends on the quantum numbers of each struck quark and antiquark (80). This observation can explain the recent analysis by Schienbein et al. (81), who find that the NuTeV measurements of nuclear structure functions obtained from neutrino charged-current reactions differ significantly from the distributions measured in deep-inelastic electron and muon scattering (**Figure 10**). This finding implies that part of the anomalous NuTeV result for $\theta_W$ could be due to the nonuniversality of nuclear antishadowing for charged and neutral currents.

The antishadowing of the nuclear structure functions as observed in deep-inelastic lepton-nucleus scattering is particularly interesting. Empirically, one finds that $R_A(x, Q^2) \equiv (F_{2A}(x, Q^2)/(A/2)F_d(x, Q^2)) > 1$ in the domain $0.1 < x < 0.2$; in other words, the measured nuclear structure function (referenced to the deuteron) is larger than the scattering on a set of $A$ independent nucleons. There are leading-twist diffractive contributions $\gamma^* N_1 \to (q\bar{q})N_1$ that arise from Reggeon exchanges in the *t* channel. For example, isospin–nonsinglet $C = +$ Reggeons contribute to the difference of proton and neutron structure functions, giving the characteristic Kuti–Weisskopf $F_{2p} - F_{2n} \sim x^{1-\alpha_R(0)} \sim x^{0.5}$ behavior at low *x*. The *x* dependence of the structure functions reflects the Regge behavior $\nu^{\alpha_R(0)}$ of the virtual Compton amplitude at fixed $Q^2$ and $t = 0$. The phase of the diffractive amplitude is determined by analyticity and crossing to be proportional to $-1 + i$ for $\alpha_R = 0.5$, which, together with the phase from the Glauber cut, leads to constructive interference of the diffractive and nondiffractive multistep nuclear amplitudes. The nuclear structure function is predicted (82) to be enhanced precisely in the domain $0.1 < x < 0.2$, in which antishadowing is empirically observed. The strength of the Reggeon amplitudes is fixed by the fits to the nucleon structure functions, so there is little model dependence. Because quarks

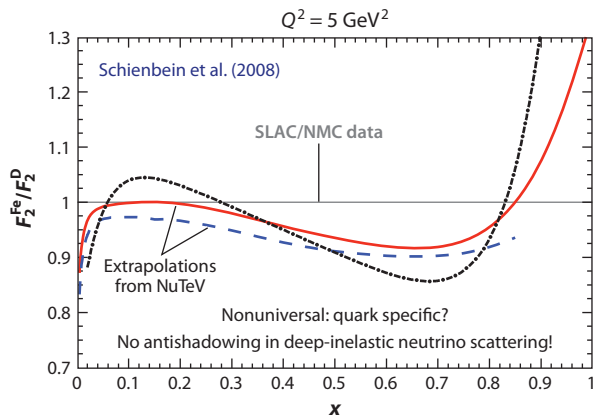

**Figure 10**

Evidence (81) that the antishadowing distribution measured in charged-current deep-inelastic neutrino-nucleus scattering differs from the antishadowing distribution measured in neutral-current deep-inelastic electron-nucleus scattering.

of different flavors couple to different Reggeons, the Reggeon amplitudes lead to the remarkable prediction that nuclear antishadowing is not universal (80); it depends on the quantum numbers of the struck quark. This scenario implies substantially different antishadowing for charged- and neutral-current reactions, thereby affecting the extraction of the weak-mixing angle $\theta_W$.

## 9.1. Dynamic Versus Static Hadronic Structure Functions

The nontrivial effects from rescattering and diffraction highlight the need for a fundamental understanding of the dynamics of hadrons in QCD at the amplitude level. This knowledge is essential for understanding phenomena such as hadronization, particularly the quantum mechanics of hadron formation, the remarkable effects of initial- and final-state interactions, the origins of diffractive phenomena and single-spin asymmetries, and manifestations of higher-twist semiexclusive hadron subprocesses.

It is usually assumed—following the intuition of the parton model—that the structure functions measured in deep-inelastic scattering can be computed in the Bjorken-scaling leading-twist limit from the absolute square of the LF wave functions, summed over all Fock states. In fact, dynamical effects, such as the Sivers spin correlation and diffractive DIS due to final-state gluon interactions, contribute to the experimentally observed deep-inelastic lepton-hadron cross sections. Diffractive events also lead to the interference of two-step and one-step processes in nuclei, which in turn, via the Gribov–Glauber theory, lead to the shadowing and the antishadowing of the deep-inelastic nuclear structure functions (80); such phenomena are not included in the LF wave functions of the nuclear eigenstate. These effects, which appear in the scattering amplitude but not within the target wave function, lead to an important distinction between dynamical and static (i.e., wave function–specific) structure functions (83).

It is therefore important to distinguish (83) static structure functions, which are computed directly from the LF wave functions of a target hadron, from the nonuniversal dynamic empirical structure functions, which take into account rescattering of the struck quark in DIS (**Figure 11**). The real wave functions of hadrons that underlie the static structure functions cannot describe diffractive deep-inelastic scattering, or single-spin asymmetries, because such phenomena involve the complex phase structure of the $\gamma^* p$ amplitude. One can augment the LF wave functions with a gauge link corresponding to an external field created by the virtual photon $q\bar{q}$ pair current (84, 85), but such a gauge link is process dependent (63), so the resulting augmented wave functions are not universal (73, 84, 86). The physics of rescattering and nuclear shadowing is not included in the nuclear LF wave functions, and a probabilistic interpretation of the nuclear DIS cross section in terms of hadron structure is thus precluded in principle, although often it can be treated as an effective approximation.

## 10. THE PROTON SPIN PROBLEM

In the naïve nonrelativistic quark model, the proton is represented as a bound state of three quarks with zero orbital angular momentum. Thus, in this approach, all of the spin $J^z$ of the proton is carried by its quark constituents. Measurements based on sum rules derived from deep-inelastic scattering measurements show that empirically, quarks carry only a small fraction of the proton's spin. Recent measurements have established that only approximately 25% of the proton's spin is carried by quarks (87).

In addition to the three valence quarks, the proton wave function contains a "sea" of nonvalence $q\bar{q}$ pairs. The dynamics induces a negative polarization of the sea quarks such that the total spin carried by quarks is reduced from the naïve 100% down to $\simeq 25\%$ of the total spin of the nucleon. This, of course, raises the question of where the remaining 75% comes from.

| Static | Dynamic |
|---|---|
| Square of target LFWFs | Modified by rescattering: ISI and FSI |
| No Wilson line | Contains Wilson line, phases |
| Probability distributions | No probabilistic interpretation |
| Process independent | Process dependent from collision |
| $T$-even observables | $T$-odd (Sivers, Boer–Mulders, etc.) |
| No shadowing, antishadowing | Shadowing, antishadowing, saturation |
| Sum rules: momentum and $J^z$ | Sum rules not proven |
| DGLAP evolution; mod. at large $x$ | DGLAP evolution |
| No diffractive DIS | Hard pomeron and odderon diffractive DIS |

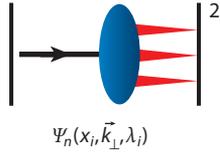
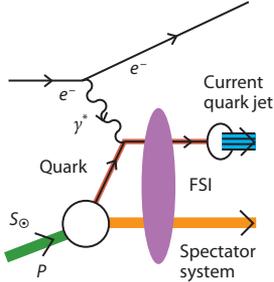

**Figure 11**

Comparison (83) between the static structure functions [determined from the light-front wave functions (LFWFs) of the hadron eigenstates] and the dynamic structure functions measured deep-inelastic lepton scattering (DIS) and other experiments. Abbreviations: DGLAP, Gribov–Lipatov–Dokshitzer–Altarelli–Parisi; FSI, final-state interaction; ISI, initial-state interaction.

The dynamical mechanism for a dramatic reduction of quark contribution to nucleon spin was in first proposed (88) in the framework of effective Lagrangians incorporating spontaneously broken chiral symmetry and a large-$N_C$ limit of QCD. The suggestion that most of the nucleon spin is probably carried by orbital angular momentum of quarks was first made within the same framework in Reference 89.

This problem can be analyzed from first principles for relativistic QCD by use of Dirac's LF quantization formalism. In this boost-invariant framework, the proton Fock state wave function describes quark and gluon constituents with nonzero orbital angular momentum. In fact, quark orbital angular momentum is required in order to have a nonzero anomalous magnetic moment, as well as the Sivers and Collins effects. Thus, one possibility is that most of the spin of the proton is effectively carried by the orbital angular momentum of the valence quarks, which is the case in AdS/QCD (90).

## 11. THE REAL PART OF THE VIRTUAL COMPTON SCATTERING AMPLITUDE

### 11.1. The $J = 0$ Fixed-Pole Contribution

At high energies, Compton scattering on an atom, $\gamma A \to \gamma A$, is dominated by the Thomson amplitude—the elastic scattering of a photon on atomic electrons. The analogous phenomenon in hadron physics is the scattering of photons on quarks $\gamma q \to \gamma q$ via either a local seagull interaction or an instantaneous fermion-exchange LF interaction. Both contributions give an energy-independent contribution to the Compton amplitude that is proportional to the charge squared of the struck quark—a contribution that does not appear in hadron-hadron scattering reactions (91). This local contribution has a real phase and is universal, providing the same contribution for real or virtual Compton scattering for any photon virtuality and skewness at fixed momentum transfer squared, $t$. The $t = 0$ limit provides an important constraint on the dependence of the nucleon mass on the quark mass through the Weisberger relation.

It is usually assumed that the imaginary part of the deeply virtual Compton scattering (DVCS) amplitude is determined at leading twist by generalized parton distributions but that the real part has an undetermined subtraction. In fact, the real part at high energies is determined by the local two-photon interactions of the quark current in the QCD LF Hamiltonian (91, 92). Because the energy dependence is independent of the momentum transfer, it is referred to as a $J = 0$ fixed pole in Regge theory. The interference of the time-like DVCS amplitude with the Bethe–Heitler amplitude leads to a charge asymmetry in $\gamma p \to \ell^+ \ell^- p$ (92, 93). Such measurements can verify that quarks carry the fundamental electromagnetic current within hadrons.

## 11.2. The Proton Radius from Muonic Hydrogen: Corrections from Virtual Compton Scattering

High-precision measurements (94) of the $\mu^- p$ Lamb shift, combined with high-order analyses in bound-state quantum electrodynamics, have led to (*a*) a determination of the proton radius that is in 5-$\sigma$ contradiction with measurements of the slope of the $G_E$ form factor at $q^2 \to 0$ and (*b*) a determination of the proton radius from the electronic Lamb shift in ordinary hydrogen. Two-photon exchange amplitudes between the muon and proton contribute an important short-distance contribution to the Lamb shift $2S - 2P$ state splitting of the muon-proton atom but are not well constrained by QCD or measurements. An anomalously large two-photon exchange contribution, including the $J = 0$ contributions, could thus provide a possible solution to the conflict between the proton radius traditionally measured in lepton-proton elastic scattering and the proton radius determined from the precision measurements of $\mu p$ atom spectroscopy.

## 12. THE ODDERON

QCD predicts an odd-$C$ exchange trajectory due to diagrams with three-gluon exchange at lowest order, a fundamental effect that has never been verified. Analyses (95, 96) predict that this trajectory has an intercept $\alpha_{\text{odderon}}(t = 0) \simeq 0$. The odderon can be measured in processes requiring odd-$C$ exchange, such as $\gamma p \to \pi^0 p'$. An even more sensitive test is to measure the difference between the charm and anticharm angular or energy distributions in $\gamma^* p \to c\bar{c} p'$. The asymmetry arises from the interference of the pomeron and odderon exchange amplitudes (97). Thus far, there is no solid evidence for the existence of the odderon—the three-gluon exchange trajectory—although it is generally regarded as a firm prediction from QCD.

## 13. ELIMINATION OF THE RENORMALIZATION-SCALE AMBIGUITY

A significant difficulty in making precise pQCD predictions is the uncertainty in determining the renormalization scale $\mu$ of the running coupling $\alpha_s(\mu^2)$. It is common to simply guess a physical scale $\mu = Q$ of order of a typical momentum transfer $Q$ in the process, then vary the scale over a range between $Q/2$ and $2Q$. This procedure is clearly problematic because the resulting fixed-order pQCD prediction depends on the choice of renormalization scheme; it can even predict negative

**QCD observables**

$$O = C[\alpha_s(\mu_0^2)] + B(\beta \log \frac{Q^2}{\mu_0^2}) + D(\frac{m_q^2}{Q^2}) + E(\frac{\Lambda_{QCD}^2}{Q^2}) + F(\frac{\Lambda_{QCD}^2}{m_Q^2}) + G(\frac{m_q^2}{m_Q^2})$$

| Scale-free conformal series | Running coupling effects | Higher twist from hadron dynamics | Intrinsic heavy quarks | Light-by-light loops |

**PMC: absorb $\beta$ terms into running coupling**

$$O = C[\alpha_s(Q^{*2})] + D(\frac{m_q^2}{Q^2}) + E(\frac{\Lambda_{QCD}^2}{Q^2}) + F(\frac{\Lambda_{QCD}^2}{m_Q^2}) + G(\frac{m_q^2}{m_Q^2})$$

**Figure 12**

Renormalization scale setting using the principle of maximal conformality (PMC). The physical result is independent of the choice of renormalization scheme and the choice of initial renormalization scale. Abbreviation: QCD, quantum chromodynamics.

QCD cross sections at NLO. If one uses the criterion that one should choose the renormalization scale to have minimum sensitivity, one gets the wrong answer in QED and even in QCD. The prediction also depends on the choice of renormalization scheme. Worse, if one tries to minimize sensitivity, the resulting renormalization scale goes to zero as the gluon jet virtuality becomes large in $e^+e^- \to q\bar{q}g$ three-jet events (98).

The running coupling in any gauge theory sums all terms involving the $\beta$ function; when the renormalization scale is set properly, all nonconformal $\beta \neq 0$ terms in a perturbative expansion arising from renormalization are summed into the running coupling (**Figure 12**). The remaining terms in the perturbative series are then identical to those of the conformal theory, namely the corresponding theory with $\beta = 0$.

The resulting scale-fixed predictions using this so-called principle of maximum conformality (PMC) are independent of the choice of renormalization scheme—a key requirement of renormalization-group invariance. In practice, the scale can be determined from the $n_f$ dependence of the NLO terms. The Brodsky–Lepage–Mackenzie (BLM)/PMC scale also determines the number of effective flavors in the $\beta$ function. The results avoid renormalon resummation and agree with QED scale setting in the Abelian limit. This is the PMC (99, 100), which underlies the BLM scale-setting method (101). Most important, the BLM/PMC method gives results that are independent of the choice of renormalization scheme, as required by the transitivity property of the renormalization group. In the case of Abelian theory, the scale is proportional to the photon virtuality and sums all vacuum polarization corrections to all orders.

Not only does the elimination of the renormalization-scheme ambiguity increase the precision of QCD tests, but it also increases the sensitivity of LHC experiments and other measurements to new physics beyond the Standard Model. The BLM/PMC method also provides scale-fixed, scheme-independent high-precision connections between observables, such as the generalized Crewther relation (102), as well as other commensurate-scale relations (103, 104).

## 14. INFRARED FEATURES OF QUANTUM CHROMODYNAMICS

It is usually assumed that the QCD coupling $\alpha_s(Q^2)$ diverges at $Q^2 = 0$, that is, "infrared slavery." In fact, determinations from lattice gauge theory, Bethe–Salpeter methods, effective charge measurements, gluon mass phenomena, and AdS/QCD lead (in their respective schemes) to a finite value of the QCD coupling in the infrared (105). Because of color confinement, the quark and gluon propagators vanish at long wavelengths: $k \ll \Lambda_{QCD}$. Consequently, the quantum-loop corrections underlying the QCD $\beta$ function decouple in the infrared, and the coupling freezes to a finite value as $Q^2 \to 0$ (106).

## 15. CONDENSATES IN QUANTUM CHROMODYNAMICS AND THE COSMOLOGICAL CONSTANT

It is conventionally assumed that the vacuum of QCD contains quark $\langle 0|q\bar{q}|0\rangle$ and gluon $\langle 0|G^{\mu\nu}G_{\mu\nu}|0\rangle$ vacuum condensates. However, as reviewed in Reference 108, the resulting vacuum energy density from QCD leads to a $10^{45}$-order-of-magnitude (or greater) discrepancy with the measured cosmological constant. In fact, Zee (107) has referred to this conflict as "one of the gravest puzzles of theoretical physics." This extraordinary contradiction between theory and cosmology has been used as an argument for the anthropic principle (108).

A possible resolution of this long-standing puzzle has recently been suggested (110); it is motivated by Bethe–Salpeter and LF analyses in which the QCD condensates are identified as "in-hadron" condensates, rather than vacuum entities, but it is consistent with the Gell-Mann–Oakes–Renner (GMOR) relation (**Figure 13**) (109). The in-hadron condensates become realized as higher Fock states of the hadron when the theory is quantized at fixed LF time: $\tau = t - z/c$.

Hadronic condensates have played an important role in QCD. Conventionally, these condensates are considered properties of the QCD vacuum and hence to be constant throughout space-time. Recently, a new perspective on the nature of QCD condensates $\langle \bar{q}q \rangle$ and $\langle G_{\mu\nu}G^{\mu\nu} \rangle$, particularly where they have spatial and temporal support, has been presented (110). A key ingredient in this approach is the use of Dirac's front form (111), namely the LF (infinite-momentum) frame, to analyze the condensates. In this formulation, the spatial support of condensates is restricted to the interior of hadrons because within the LF vacuum is an empty Fock state. Thus, condensates arise due to the interactions between quarks and gluons, which are confined within hadrons.

The radical new idea is to apply this revised view of QCD condensates to the long-standing problem of the cosmological constant in quantum field theory. This proposal is a paradigm shift, and many theorists (including one of the authors of this review, Marek Karliner) view it with considerable skepticism. This skepticism is mostly due to intuition coming from more traditional approaches to the problem of condensates in quantum field theory, such as effective Lagrangians

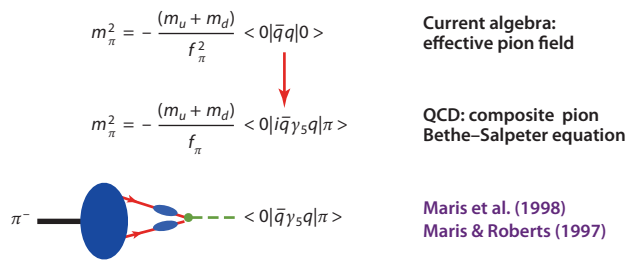

**Figure 13**

In-hadron condensates and the revised GMOR (Gell-Mann–Oakes–Renner) relation derived in quantum chromodynamics (QCD) in References 114 and 115 by use of the Bethe–Salpeter formalism. Vacuum-to-vacuum matrix elements do not appear.

and lattice gauge theory. Nevertheless, the new idea is very intriguing and it is based on well-tested principles. The stakes are very high, so in the following paragraphs we briefly review the details of this proposal, hoping that this review will lead to a constructive discussion that may result in a new understanding of the problem.

When one makes a measurement in hadron physics, such as deep-inelastic lepton-proton scattering, one probes hadron's constituents consistent with causality—at a given LF time, not at instant time. Similarly, when one makes observations in cosmology, information is obtained within the causal horizon; that is, it is consistent with the finite speed of light.

Physical eigenstates are built from operators acting on the vacuum. It is therefore important to distinguish two very different concepts of the vacuum in quantum field theories such as QED and QCD. The conventional instant-form vacuum is a state defined at the same time $t$ at all spatial points in the universe. In contrast, the front-form vacuum senses only phenomena that are causally connected, that is, within the observer's light-cone. The instant-form vacuum is defined as the lowest-energy eigenstate of the instant-form Hamiltonian. For example, the instant-form vacuum in QED is saturated with quantum loops of leptons and photons. In calculations of physical processes, one must then normal-order the vacuum and divide the $S$-matrix elements by the disconnected vacuum loops. In contrast, the front-form vacuum at fixed LF time is defined as the lowest-mass eigenstate of LF Hamiltonian quantized at fixed $\tau = t - z/c$. The LF vacuum is remarkably simple in LF quantization because of the restriction that $k^+ \geq 0$. For example, in QED, vacuum graphs, such as $e^+e^-\gamma$, associated with the zero-point energy do not arise. In the Higgs theory, the usual Higgs vacuum expectation value is replaced with a $k^+ = 0$ zero mode (112); however, the resulting phenomenology is identical to the standard analysis. The LF vacuum thus coincides with the vacuum of the free LF Hamiltonian (up to possible zero modes set by boundary conditions). The front-form vacuum is causal and Lorentz invariant, whereas the instant-form vacuum is acausal and depends on the observer's Lorentz frame. In the LF theory, the QCD condensate physics is replaced by the dynamics of higher nonvalence Fock states, as shown by Casher & Susskind (113). In particular, chiral symmetry is broken in a limited domain of size $1/m_\pi$, which is analogous to the limited physical extent of superconductor phases.

The cosmological constant measures the matrix element of the energy momentum tensor $T^{\mu\nu}$ in the background universe. It corresponds to the measurement of the gravitational interactions of a probe of finite mass; it senses only the causally connected domain within the light-cone of the observer. If the universe is empty, the appropriate vacuum state is therefore the LF vacuum because it is causal. One automatically obtains a vanishing cosmological constant from the LF vacuum. Thus, as argued in Reference 110, the many-orders-of-magnitude conflict between QCD and the observed value of the cosmological condensate is removed, and a new perspective on the nature of quark and gluon condensates in QCD is thereby obtained.

However, in the LF analysis one finds that the spatial support of QCD condensates is restricted to the interior of hadrons, physics that arises due to the interactions between confined quarks and gluons. The condensate physics normally associated with the instant-form vacuum is replaced by the dynamics of higher nonvalence Fock states, as shown by Casher & Susskind (113) in the context of the infinite momentum method. In particular, chiral symmetry is broken in a limited domain of size $1/m_\pi$, in analogy to the limited physical extent of superconductor phases. This novel description of chiral symmetry breaking in terms of in-hadron condensates has also been observed in Bethe–Salpeter studies (114, 115). The usual argument for a quark vacuum condensate is the GMOR formula: $m_\pi^2 = -2m_q \langle 0|\bar{q}q|0\rangle/f_\pi^2$. However, in the Bethe–Salpeter and LF formalisms, in which the pion is a $q\bar{q}$ bound state, the GMOR relation is replaced by $m_\pi^2 = -2m_q \langle 0|\bar{q}\gamma_5 q|\pi\rangle/f_\pi$, where $\rho\pi \equiv -\langle 0|\bar{q}\gamma_5 q|\pi\rangle$ represents a pion decay constant via an elementary pseudoscalar current.

The result is independent of the renormalization scale. In the LF formalism, this matrix element derives from the $|q\bar{q}\rangle$ Fock state of the pion with parallel spin projections $S^z = \pm 1$ and $L^z = \mp 1$, which couple by quark spin-flip to the usual $|q\bar{q}\rangle$ $S^z = 0$, $L^z = 0$ Fock state via the running quark mass.

This new perspective might explain the results of studies (116–118) that find no significant signal for the vacuum gluon condensate. Thus, one finds in-hadron condensates replacing vacuum condensates: The $\langle 0|\overline{qq}|0\rangle$ vacuum condensate that appears in the GMOR formula is replaced by the $\langle 0|\bar{q}\gamma_5 q|\pi\rangle$ pion decay constant.

AdS/QCD also provides a description of chiral symmetry breaking by using the propagation of a scalar field $X(z)$ to represent the dynamical running quark mass. In the hard wall model, the solution has the form (119, 120) $X(z) = a_1 z + a_2 z^3$, where $a_1$ is proportional to the current-quark mass and $a_2$ represents a $\bar{q}q$ expectation value in the confined domain. The spatial variation of the scalar field $X$ within the hadronic domain is thus different from a uniform vacuum condensate. The coefficient $a_2$ scales as $\Lambda_{QCD}^3$ and is the analog of $\langle\bar{q}q\rangle$; however, because the quark is a color nonsinglet, the propagation of $X(z)$, and thus the domain of the quark condensate, is limited to the region of color confinement. Furthermore, the effect of the $a_2$ term varies within the hadron, as is characteristic of an in-hadron condensate.

## 16. NONANALYTIC FEATURES OF THE PROTON SIZE FROM CHIRAL PERTURBATION THEORY

Chiral perturbation theory is based on an effective theory of elementary pions and nucleons, which is expected to be a valid approximation to QCD at low momentum transfer. A remarkable prediction of this approach is the nonanalytic behavior of the radii derived from the derivative of the isovector pion and nucleon form factors at $q^2 = 0$. For example, the chiral model predicts that the square of the radius derived from the slope of the Pauli form factor diverges as $(1/m_\pi)$ at $m_\pi \to 0$ (121). These phenomena are due to the cut of the time-like form factor at $q^2 = 4m_\pi^2 \to 0$. There are still open questions as to whether such nonanalytic phenomena can be demonstrated directly from QCD. This phenomenon is not apparent in the AdS/QCD framework discussed in Section 18, even at zero quark mass, because the quarks of the higher Fock states are contained within the confinement radius.

## 17. PHOTON-TO-MESON TRANSITION FORM FACTORS

The photon-to-meson transition form factors (TFFs) $F_{M\gamma}(Q^2)$ measured in $\gamma\gamma^* \to M$ reactions have attracted intense experimental and theoretical interest. The pion TFF between a photon and a pion measured in the $e^-e^- \to e^-e^-\pi^0$ process, with one tagged electron, is the simplest bound-state process in QCD. It can be predicted from first principles in the asymptotic $Q^2 \to \infty$ limit (16). More generally, the pion TFF at large $Q^2$ can be calculated at leading twist as a convolution of a perturbative hard-scattering amplitude $T_H(\gamma\gamma^* \to q\bar{q})$ and a gauge-invariant meson distribution amplitude that incorporates the nonperturbative dynamics of the QCD bound state (16).

The BaBar Collaboration has reported measurements of the TFFs from the $\gamma^*\gamma \to M$ process for the $\pi^0$ (122), $\eta$, and $\eta'$ (123, 124) pseudoscalar mesons for a momentum transfer range much larger than that used in previous measurements (125, 126). Surprisingly, the BaBar data for the $\pi^0 - \gamma$ TFF exhibit a rapid growth for $Q^2 > 15\, \text{GeV}^2$, which is unexpected from QCD predictions. In contrast, the data for the $\eta - \gamma$ and $\eta' - \gamma$ TFFs agree with data from previous experiments and theoretical predictions. Many theoretical studies have been devoted to explaining BaBar's experimental results (127–138).

The pion TFF $F_{\pi\gamma}(Q^2)$ can be computed from first principles in QCD. To LO in $\alpha_s(Q^2)$ and to leading twist, the result is (16) ($Q^2 = -q^2 > 0$)

$$Q^2 F_{\pi\gamma}(Q^2) = \frac{4}{\sqrt{3}} \int_0^1 dx \frac{\phi(x, \bar{x}Q)}{\bar{x}} \left[1 + O\left(\alpha_s, \frac{m^2}{Q^2}\right)\right], \qquad 6.$$

where $x$ is the longitudinal momentum fraction of the quark struck by the virtual photon in the hard-scattering process and $\bar{x} = 1 - x$ is the longitudinal momentum fraction of the spectator quark. The pion distribution amplitude $\phi(x, Q)$ in the LF formalism (16) is the integral of the valence $q\bar{q}$ LF wave function in light-cone gauge $A^+ = 0$,

$$\phi(x, Q) = \int_0^{Q^2} \frac{d^2\mathbf{k}_\perp}{16\pi^3} \psi_{q\bar{q}/\pi}(x, \mathbf{k}_\perp), \qquad 7.$$

and has the asymptotic form (16) $\phi(x, Q \to \infty) = \sqrt{3} f_\pi x(1 - x)$. Thus, the LO QCD result for the TFF at the asymptotic limit is obtained (16): $Q^2 F_{\pi\gamma}(Q^2 \to \infty) = 2 f_\pi$.

The LF holographic methods described in Section 18 can be used in the analysis of the two-photon processes $\gamma\gamma^* \to M$ (139). A simple analytical expression for the pion TFF can be obtained from the "soft-wall" holographic model described in Section 18,

$$Q^2 F_{\pi\gamma}(Q^2) = \frac{4}{\sqrt{3}} \int_0^1 dx \frac{\phi(x)}{1-x} \left[1 - \exp\left(-\frac{(1-x)P_{q\bar{q}}Q^2}{4\pi^2 f_\pi^2 x}\right)\right], \qquad 8.$$

where $\phi(x) = \sqrt{3} f_\pi x(1-x)$ is the asymptotic QCD distribution (here $f_\pi$ is the pion decay constant and $P_{q\bar{q}}$ is the probability for the valence state). Remarkably, the holographic result for the pion TFF given by Equation 8 for $P_{q\bar{q}} = 1$ is identical to the results for the pion TFF obtained with the exponential LF wave function model of Musatov & Radyushkin (140), consistent with the LO QCD result (16).

By taking $P_{q\bar{q}} = 0.5$ in Equation 8, one obtains a result that agrees with the Adler–Bell–Jackiw (141) anomaly result, which agrees within a few percent with the observed value obtained from the decay $\pi^0 \to \gamma\gamma$. This agreement suggests that the contribution from higher Fock states vanishes at $Q = 0$ in this simple holographic confining model. Thus, Equation 8 represents a description of the pion TFF that encompasses the low-energy nonperturbative and high-energy hard domains but includes only the asymptotic distribution amplitude of the $q\bar{q}$ component of the pion wave function at all scales. The results from Equation 8 are shown in **Figure 14a** for $Q^2 F_{\pi\gamma}(Q^2)$. The calculations (139) agree reasonably well with the experimental data at low- and medium-$Q^2$ regions ($Q^2 < 10$ GeV$^2$) but disagree with BaBar's high-$Q^2$ data.

The $\eta$ and $\eta'$ mesons result from the mixing of the neutral states $\eta_8$ and $\eta_1$ of the $SU(3)_F$ quark model. The TFFs for the $\eta$ and $\eta'$ mesons have the same expression as the pion TFF, except for an overall multiplying factor of $c_P = 1$, $1/\sqrt{3}$, or $(2\sqrt{2}/\sqrt{3})$ for the $\pi^0$, $\eta_8$, or $\eta_1$, respectively (139). The results for the $\eta$ and $\eta'$ TFFs are shown in **Figure 14b,c** for $Q^2 F_{M\gamma}(Q^2)$. The calculations agree very well with available experimental data over a wide range of $Q^2$. Furthermore, the predictions for the $\eta$ and $\eta'$ TFFs remain largely unchanged if other mixing schemes (142, 143) are used in the calculation.

The rapid growth of the high-$Q^2$ data for the pion-photon TFF reported by the BaBar Collaboration is difficult to explain within the current framework of QCD. If the BaBar data for the meson-photon TFF are confirmed, they could indicate physics beyond the Standard Model, such as a weakly coupled elementary $C = +$ axial vector or a pseudoscalar $z^0$ in the few-GeV domain, an elementary field that would provide the coupling $\gamma^*\gamma \to z^0 \to \pi^0$ at leading twist.

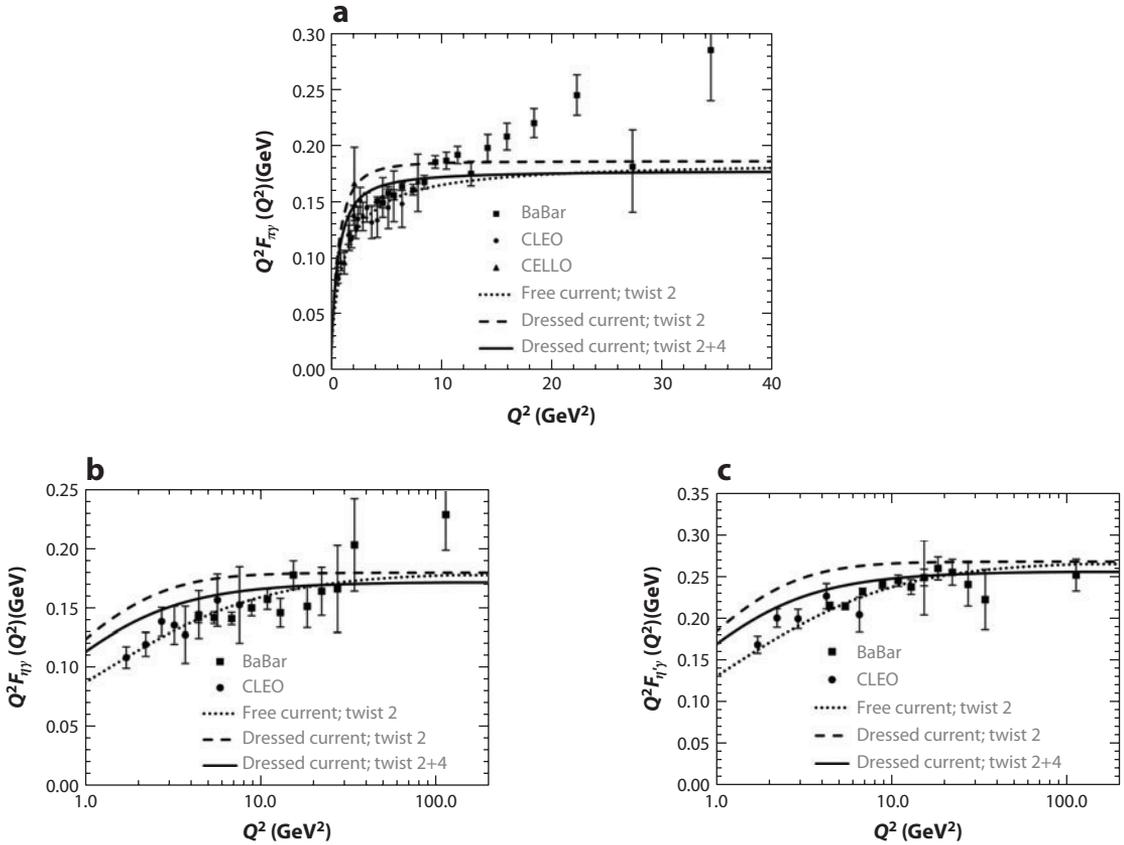

**Figure 14**

(*a*) The $\gamma\gamma^* \to \pi^0$ photon-to-pion transition form factor $Q^2 F_{\pi\gamma}(Q^2)$. The dashed and solid curves include the effects of using a confined electromagnetic current for twist-2 and twist-2+4, respectively. (*b*) Same for the $\gamma\gamma^* \to \eta$ transition form factor. (*c*) Same for the $\gamma\gamma^* \to \eta'$ transition form factor. The data are from References 122, 125, and 126.

The analysis presented here thus indicates the importance of obtaining additional measurements of the pion-photon TFF at large $Q^2$.

## 18. ANTI–DE SITTER/QUANTUM CHROMODYNAMICS AND LIGHT-FRONT HOLOGRAPHY

A long-sought goal in hadron physics is to find a simple analytic first approximation to QCD that is analogous to the Schrödinger-Coulomb equation of atomic physics. This problem is particularly challenging because the formalism must be relativistic, color confining, and consistent with chiral symmetry. de Téramond & Brodsky (144) have shown that the gauge/gravity duality leads to a simple analytical and phenomenologically compelling nonperturbative approximation to the full LF QCD Hamiltonian: LF holography (144). LF holography is one of the most remarkable features of the AdS/CFT (conformal field theory) correspondence (145). In particular, the soft-wall AdS/QCD model, modified by a positive-sign dilaton metric, leads to a simple Schrödinger-like LF wave equation and a remarkable one-parameter description of nonperturbative hadron dynamics

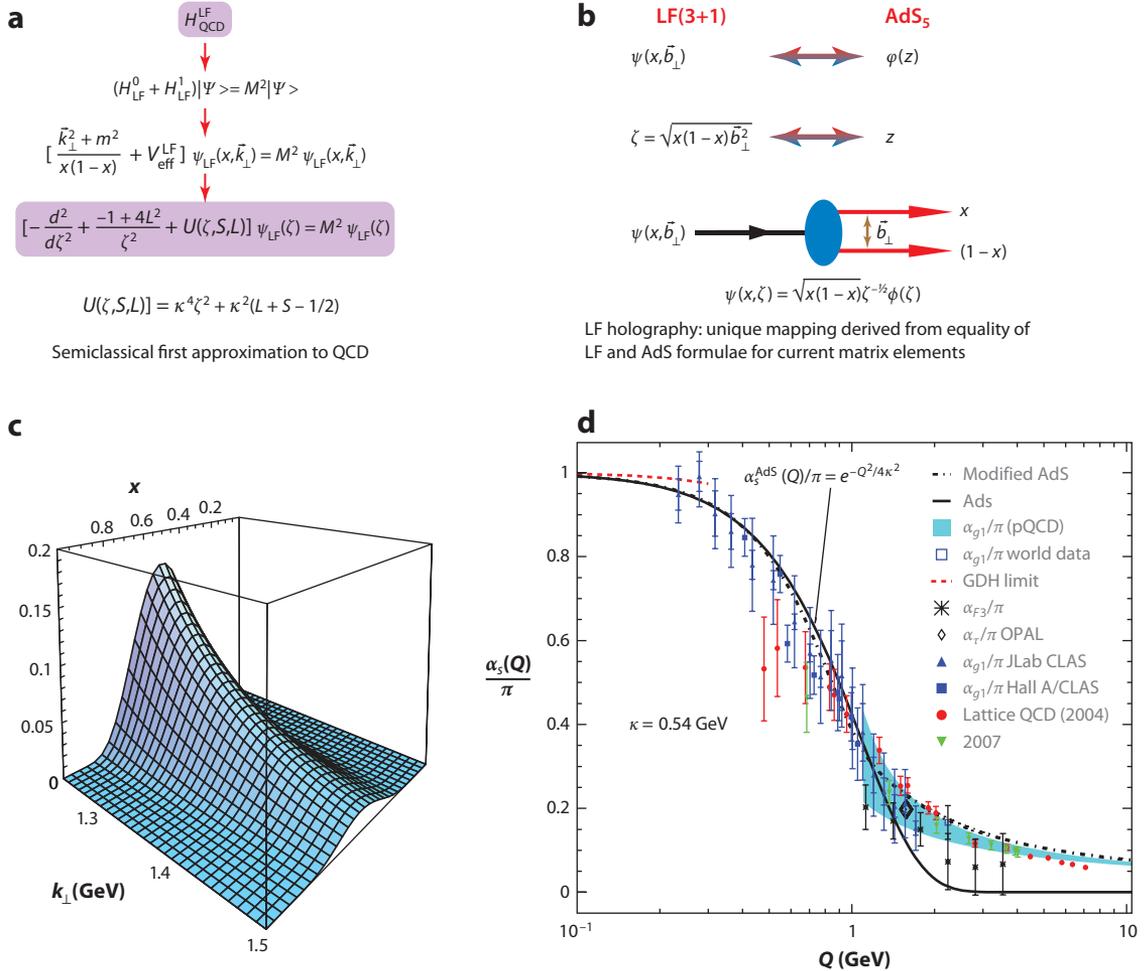

**Figure 15**

(*a*) Reduction of the light-front (LF) Hamiltonian to an effective LF Schrödinger equation for mesons. (*b*) Mapping of the fifth-dimension coordinate $z$ to the invariant LF separation variable $\zeta$. (*c*) The anti–de Sitter/quantum chromodynamics (AdS/QCD)–LF holography prediction for the pion's valence LF wave function $\psi(x, k_\perp)$. (*d*) The running coupling predicted by AdS/QCD, normalized to $\alpha_s/\pi = 1$, compared with the effective charge defined from the Bjorken sum rule. Abbreviation: pQCD, perturbative quantum chromodynamics. Reproduced from Reference 105.

(144, 146, 147). The model predicts a zero-mass pion for massless quarks and a Regge spectrum of linear trajectories with the same slope in the (leading) orbital angular momentum $L$ of the hadrons and their radial quantum number $N$.

LF holography maps the amplitudes that are functions of the fifth-dimension variable $z$ of AdS space to a corresponding hadron theory quantized on the LF (139, 144). The resulting Lorentz-invariant relativistic LF wave equations are functions of an invariant impact variable $\zeta$ that measures the separation between the quark and gluonic constituents within the hadron at equal LF time (**Figure 15***c*). Remarkably, the AdS equations correspond to the kinetic energy terms of the partons inside a hadron, whereas the interaction terms build confinement and correspond to the truncation of AdS space in an effective dual gravity approximation (144). The identification of the orbital angular momentum of the constituents is a key element in our description of the internal structure of hadrons using holographic principles, given that hadrons with the same quark content, but different orbital angular momenta, have different masses.

The result is a semiclassical frame-independent first approximation to the spectra and LF wave functions of meson and baryon light quark bound states that, in turn, predicts the behavior of the pion and nucleon form factors. The hadron eigenstates generally have components with different orbital angular momentum; for example, the proton eigenstate in AdS/QCD with massless quarks has $L^z = 0$ and $L^z = 1$ LF Fock components with equal probability. Thus, in AdS/QCD the spin of the proton is carried by the quark orbital angular momentum: $J^z = \langle L^z \rangle = \pm 1/2$ because $\langle \sum S_q^z \rangle = 0$ (90), which helps to explain the spin crisis (discussed in more detail in Section 10).

The AdS/QCD soft-wall model also predicts the form of the nonperturbative effective coupling $\alpha_s^{\text{AdS}}(Q)$ (**Figure 15d**) and its $\beta$ function, in excellent agreement with JLab measurements (105). The AdS/QCD LF wave functions have also led to a proposal for computing the hadronization of quark and gluon jets at the amplitude level (149).

In general, the QCD Hamiltonian can be systematically reduced to an effective equation in acting on the valence Fock state. **Figure 15a** illustrates this reduction for mesons. The kinetic energy contains a term, $L^2/\zeta^2$, that is analogous to $\ell(\ell+1)/r^2$ in nonrelativistic theory, where the invariant $\zeta^2 = x(1-x)b_\perp^2$ is conjugate to the $q\bar{q}$ invariant mass $k_\perp^2/x(1-x)$. It plays the role of the radial variable $r$. Here, $L = L^z$ is the projection of the orbital angular momentum appearing in the $\zeta, \phi$ basis. In QCD, the interaction $U$ couples the valence state to all Fock states. The AdS/QCD model has the same structure as the reduced form of the LF Hamiltonian, but it also specifies the confining potential as $U(\zeta, S, L) = \kappa^4 \zeta^2 + \kappa^2(L + S - 1/2)$. This correspondence, plus the fact that one can match the AdS/QCD formulae for elastic electromagnetic (106) and gravitational form factors (148) to the LF Drell–Yan–West formula, is the basis of LF holography. The light quark meson and baryon spectroscopy is well described by taking the mass parameter $\kappa \simeq 0.5$ GeV. The linear trajectories in $M_H^2(n, L)$ have the same slope in $L$ and $n$, the radial quantum number. The corresponding LF wave functions are functions of the off-shell invariant mass. AdS/QCD, together with LF holography (144), thus provides a simple Lorentz-invariant, color-confining approximation to QCD that is successful in accounting for light quark meson and baryon spectroscopy, as well as their LF wave functions. The semiclassical approximation to LF QCD described in this section is expected to break down at short distances where hard gluon exchange and quantum corrections become important. The model can be systematically improved by use of Lippmann–Schwinger methods (150) or by use of the AdS/QCD orthonormal basis to diagonalize the LF Hamiltonian.[1] One can also improve the semiclassical approximation by introducing nonzero quark masses and short-range Coulomb corrections, thereby extending the predictions of the model to the dynamics and spectra of heavy and heavy-light quark systems (151). For a recent review of the AdS/QCD approach, see Reference 152.

## 19. SUBLIMATED GLUONS

An important feature of AdS/QCD is its prediction of the running coupling constant in the infrared domain: $\alpha_s^{\text{AdS/QCD}}(Q^2) \propto e^{-Q^2/4\kappa^2}$ (105). The Gaussian falloff of the AdS/QCD prediction for $\alpha_s^{\text{AdS/QCD}}$ is clearly incorrect in the short-distance domain $Q^2 > 4\kappa^2 \simeq 1$ GeV$^2$, where the asymptotic freedom property of QCD becomes evident, as seen from the falloff of the measured $g_1$ effective charge $\alpha_s^{g_1}(Q^2)$ (153). We can interpret the breakdown of the AdS/QCD prediction at hard scales as evidence of the appearance of dynamical gluon degrees of freedom in the $Q^2 > \text{GeV}^2$ domain. However, gluons with smaller virtuality are sublimated in terms of the effective confining potential.

---

[1]We thank J. Vary for helpful conversations on this issue.

An essential prediction of QCD is the existence of color-octet spin-one gluon quanta. If one quantizes QCD by using LF quantization and a $A^+ = 0$ light-cone gauge, then the gluon quanta have positive metric and physical polarization: $S^z = \pm 1$. Gluon jets are clearly observed in hard QCD processes, such as the three-jet events in electron-positron annihilation, $e^+e^- \to q\bar{q}g$ (154). Empirically, the rapidity plateau of a gluon jet is higher than that of a quark jet by the factor $\frac{C_A}{C_F} = \frac{9}{4}$, as predicted at LO in QCD. The existence of asymptotic freedom at large $Q^2$: $\alpha_s(Q^2) \simeq 4\pi/\beta_0 \log \frac{Q^2}{\Lambda^2_{\text{QCD}}}$, the DGLAP evolution of structure functions, and the ERBL (Efremov–Radyushkin–Brodsky–Lepage) evolution of distribution amplitudes are all based on the existence of hard gluons. However, empirical evidence confirming gluonic degrees of freedom at small virtualities is lacking. For example:

1. Finding clear evidence for $gg$ and $ggg$ gluonium bound states has been difficult, even in the "gluon factory" reaction $J/\psi \to \gamma g g$ (155, 156). Similarly, there is no clear experimental evidence for $q\bar{q}g$ hybrid states (155).[2]
2. One would normally expect gluon exchange to dominate large-angle, elastic scattering, exclusive hadron-hadron scattering reactions. In fact, as we discussed in Section 4, two-body scattering amplitudes at fixed $\theta_{\text{cm}}$ are dominated by quark exchange and interchange amplitudes (26) rather than by gluon exchange contributions.
3. Experiments find that the $J/\psi \to \rho\pi$ is the largest two-body mode, whereas $\psi'$ almost never decays to $\rho\pi$. The infamous $J/\psi \to \rho\pi$ puzzle shows that the usual OZI picture of $c\bar{c}$ annihilation to three gluons, each with virtuality $q^2 \simeq M^2_\psi/9$, is incorrect. However, this puzzling decay pattern is consistent with a mechanism (3) in which the $c\bar{c}$ of the quarkonium state flows into the $|q\bar{q}c\bar{c}\rangle$ Fock state of one of the final-state mesons. The suppression of $\psi' \to \rho\pi$ is then due to the node in the excited quarkonium radial wave function.
4. AdS/QCD makes the remarkable prediction that the proton spin is carried by the orbital angular momentum $\langle J^z \rangle = \langle L^z \rangle = \pm 1/2$. However, because of the $SU(6)$ weights, we have

$$\langle L^z \rangle_u = \frac{4}{3} \times \frac{1}{2} = \frac{2}{3} \qquad 9.$$

and

$$\langle L^z \rangle_d = -\frac{1}{3} \times \frac{1}{2} = -\frac{1}{6}. \qquad 10.$$

In other words, the orbital angular momentum carried by the two up quarks and one down quark in a polarized proton with $J^z = +1/2$ is very different. The ratio is $4:(-1)$. This result provides a new way to understand the large Sivers effect for the struck $u$ quarks, compared with the small Sivers effect for the $d$ quark, because it depends on orbital angular momentum.

These striking phenomenological features are consistent with the AdS/QCD prediction that gluon quanta with virtuality below $Q^2 \simeq 1\,\text{GeV}^2$ are physically absent. Thus, at low $Q^2$ the physical effects of gluons are evidently sublimated and replaced by an effective potential that confines quarks. The AdS/QCD scenario is consistent with string descriptions of confinement and the Isgur–Paton flux tube model (6). Interestingly enough, gluon sublimation might be the first step toward resolving the long-standing challenge of deriving the naïve quark model from QCD. In the quark model, there are no gluons, yet it is unaccountably successful in describing the low-lying spectrum of QCD.

---

[2]There is, however, a clear evidence for glueballs from lattice gluodynamics, namely QCD with $N_f = 0$.

## 20. CONCLUSIONS

In this review, we highlight many examples in which experimental results in hadron physics do not agree with conventional expectations. On the one hand, some of these anomalies and puzzles, such as the BaBar measurements of the photon-to-pion TFF, may challenge the assumption that QCD is the correct fundamental theory of the strong and nuclear interactions. On the other hand, these striking examples of novel physics more likely suggest that we have not fully uncovered the remarkable features and subtleties of the theory. Thus, in many cases, surprising new perspectives for QCD hadronic and nuclear physics have emerged. We point out numerous areas wherein often-used procedures in QCD and hadron physics have been challenged. These include the following conventional assumptions.

1. The structure function of a hadron reflects only the physics of the wave function of the hadron and thus must be process independent. In fact, the observed structure functions are sensitive to rescattering processes at leading twist, which are process dependent.
2. Antishadowing is a property of the nuclear wave function and is thus process independent. In fact, as the NuTeV data show, each quark may have its own antishadowing distribution.
3. Initial-state and final-state interactions are always power-law suppressed and process independent. This hypothesis is contradicted by the Sivers effect in SIDIS and the breakdown of the Lam–Tung relation in Drell–Yan reactions.
4. High–transverse momentum hadrons always arise only from jet fragmentation. In fact, there is a significant probability that high-$p_T$ hadrons arise from hard color-transparent subprocesses. As we discuss above, direct higher-twist processes wherein the hadron wave function appears in the subprocess matrix element can explain anomalies in the fixed-$x_T$ cross section and the remarkable baryon anomaly, the large proton-to-pion ratio observed in heavy-ion collisions at RHIC.
5. The renormalization scale in QCD cannot be fixed and can only be guessed to minimize sensitivity. In fact, it can be fixed at each order in perturbation theory in a scheme-independent way that agrees with the conventional QED procedure.
6. QCD condensates must be properties of the vacuum. As we discuss above, contrary results are obtained in Bethe–Salpeter and LF analyses. The conflict with the cosmological constant highlights the need to distinguish different concepts of the vacuum obtained from the usual instant form versus the causal LF definition.
7. Infrared slavery: The QCD running coupling must diverge at long distances. This is not correct in LF holographic QCD, nor is it true if one defines the QCD coupling through an effective charge defined from experiment.
8. Nuclei can be regarded as composites of color-singlet nucleons. In fact, QCD predicts hidden-color configurations of the quarks, which can dominate short-distance nuclear reactions.
9. The real part of DVCS is an arbitrary subtraction term. In fact, local four-point photon-quark scattering can lead to a novel amplitude that is constant in energy and independent of the photons' virtuality at fixed $t$.
10. Heavy quark thresholds cause minimal effects. In fact, the charm and strangeness thresholds can lead to unexpectedly large competing amplitudes and striking polarization effects, such as the remarkable spin-spin correlations observed in elastic *pp* scattering and, at large angles, the breakdown of pQCD color transparency.
11. Gluon degrees of freedom should be manifest at all scales. In fact, the effects of soft gluons may well be sublimated in favor of the QCD confinement potential.
12. Orbital angular momentum effects are negligible. In fact, in the LF framework the hadron eigensolutions for the light quarks have orbital components that are comparable in strength to the $L = 0$ components.
13. The heavy quark sea arises only from gluon splitting and is thus confined to the low-*x* domain. In fact, QCD predicts contributions where the heavy quarks are multiconnected to the valence quarks and thus appear at high *x*.

## DISCLOSURE STATEMENT

The authors are not aware of any affiliations, memberships, funding, or financial holdings that might be perceived as affecting the objectivity of this review.